\documentclass{kapproc} 
\usepackage{procps} 

\usepackage[dvips]{graphicx}

\upperandlowercase

\nochapequationnumber 

\let\footnote\savefootnote

\let\footnotetext\savefootnotetext

\normallatexbib 

\begin{document}



\articletitle
[]
{The QGP Discovered at RHIC}

\author{M. Gyulassy}




\affil{Physics Department, Columbia University, New York, USA}
\email{gyulassy@nt3.phys.columbia.edu}





--------------

\begin{abstract}
Three empirical lines of evidence, $({\rm \bf P_{QCD}, pQCD, dA)}$, from RHIC
have converged and point to the discovery of a strongly coupled Quark Gluon Plasma.
The evidence includes (1) bulk collective elliptic flow and (2) jet quenching and mono-jet
production, observed in Au+Au collisions at 200 AGeV, and (3) a critical
control experiment using D+Au at 200 AGeV. 
\end{abstract}


\section{The Theoretical QGP}
The Standard Model of strong interactions predicts the existence of a new phase of matter,
called a Quark Gluon Plasma (QGP), in which the quark and gluon degrees of freedom
normally confined within hadrons are mostly liberated. Lattice QCD calculations
show that there is a rapid rise of the entropy density, $\sigma(T)$,
 of matter when the temperature
reaches $T\approx T_c\sim 160$ MeV. Beyond $T_c$
the effective number of degrees of freedom, $n(T)$,
saturates near the number of quark and gluon helicity states 
$n_{QCD}=8_c\times 2_s + \frac{7}{8}\times 3_c\times N_f\times 2_s\times 2_{q\bar{q}}\approx
37$. The entropy density $\sigma(T)=dP/dT=(\epsilon+P)/T\propto n(T)T^3$ 
approaches the Stefan Boltzmann limit
$4P_{SB}(T)/3T$.
The transition region is a smooth crossover when dynamical
quarks are taken into account, but the width of the transition region remains relatively 
narrow, $\Delta T_c/T_c \sim 0.1$ \cite{Fodor:2004nz}-\cite{Rischke:2003mt}.

\begin{figure}[h]
\centering
\vspace*{-0.6cm}
\includegraphics[height=0.35\textheight,width=1.0\textwidth,clip]{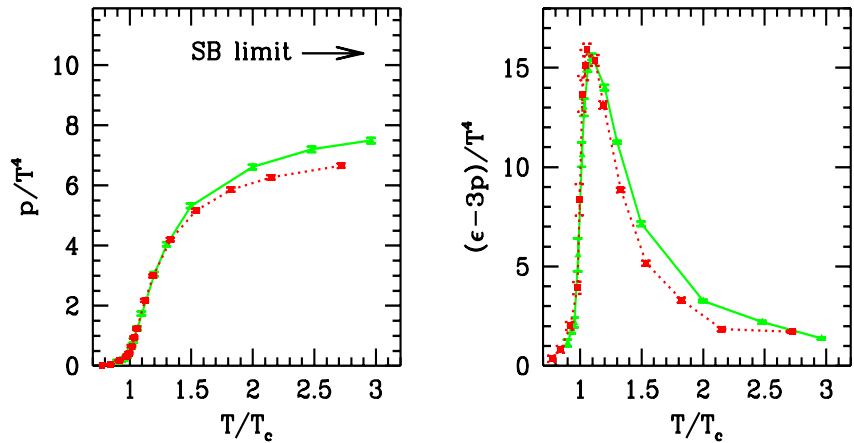}
\vspace*{-0.7cm}
\caption{A recent Lattice QCD calculation \protect{\cite{Fodor:2004nz}} 
of the pressure, $P(T)/T^4$, and a measure of the deviation
from the ideal Stefan-Boltzmann limit $(\epsilon(T)-3 P(T))/T^4$.
Note that the scale on both graphs has not been corrected for finite lattice volume
effects: see \protect{\cite{Allton:2003vx}} for discussion. }
\label{qgpfig1}
\end{figure}

The rapid rise of the entropy was predicted long before QCD 
by Hagedorn due to the observed exponential rise of the
number of hadron resonances~\cite{Karsch:2003zq}. However, the saturation of
the number of degrees of freedom near $n_{QCD}$ is a unique feature of QCD. 
Even though the entropy density approaches the ideal, weakly interacting
plasma limit, 
lattice calculations of correlators show that the 
QGP is far from ideal below $3T_c$. 
The nonideal nature of this strongly coupled QGP is also seen from the 
deviation of the pressure, $P(T)$, and energy density $\epsilon(T)$
from the Stefan Boltzmann limit 
as shown in Fig.(\ref{qgpfig1}) from \cite{Fodor:2004nz}. 

\begin{figure}[h]
\centering
\vspace*{-1cm}
\includegraphics[height=0.35\textheight,width=0.45\textwidth,clip]{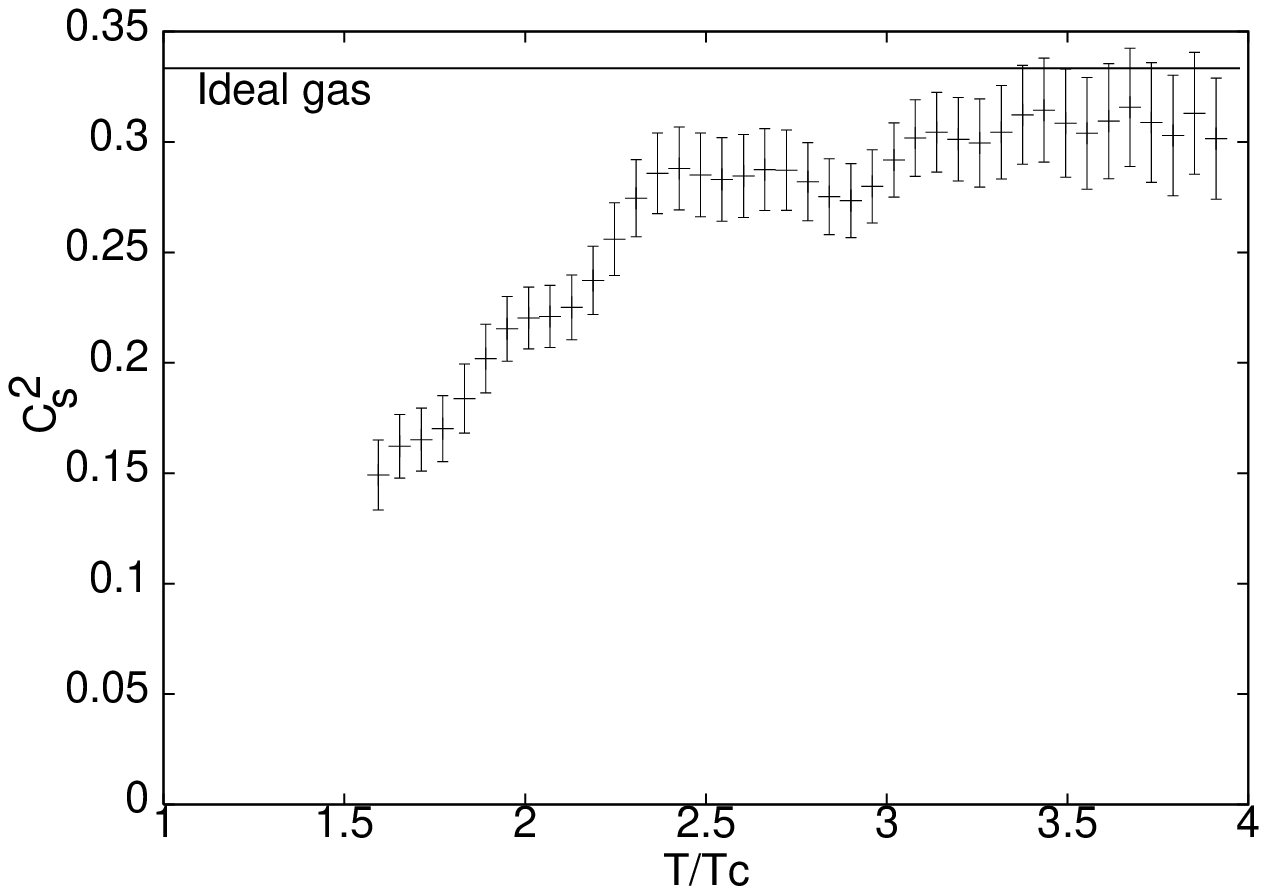}
\hfill
\includegraphics[height=0.45\textheight,width=0.45\textwidth,clip]{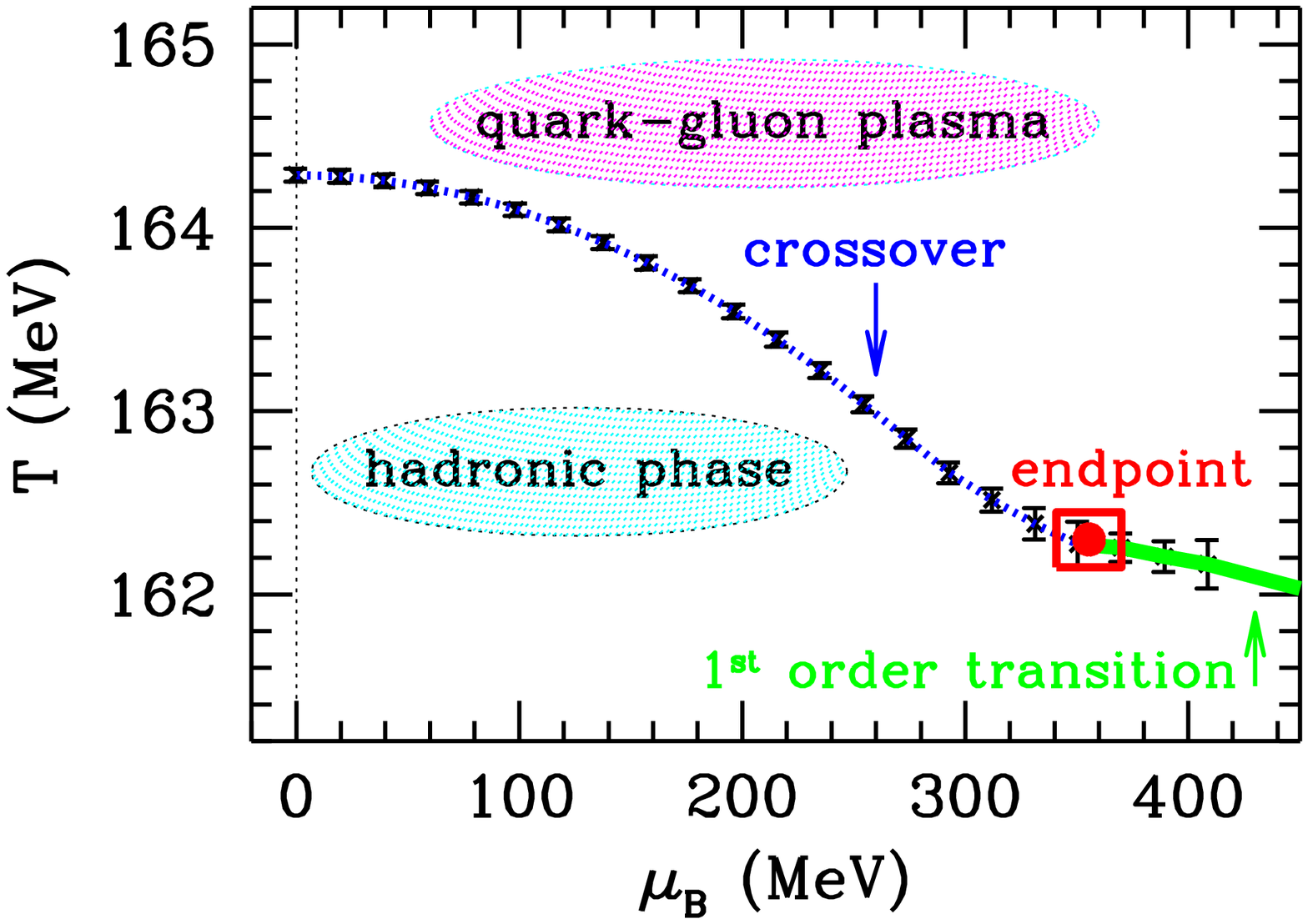}
\caption{Important features of the QGP equation of state.
The speed of sound \protect{\cite{Gupta:2003be}} 
$c_s^2=d\epsilon/dP$ drops below $1/3$ for $T<2T_c \approx300$ MeV.
Right panel shows a current estimate of the location of the 
tricritical point at finite baryon density \protect{\cite{Fodor:2004nz}} 
}
\label{qgpfig2}
\end{figure}

The equation of state of the QGP, $P_{QCD}(T)$,  is  
the bulk thermodynamic property that can be investigated experimentally
via ``barometric'' observables. 
A measure of its  stiffness is given by  the  speed of 
sound squared, $c_s^2=dP/d\epsilon=d\log T/d\log\sigma=(3+d\log n/d\log T)^{-1}$ 
shown in  Fig.(\ref{qgpfig2}).
Note that $c_s^2$  drops rapidly below $1/3$ as the effective number of degrees
of freedom drops 
when $T$ approaches $T_c$. This softening of the QGP equation of state near $T_c$
is a key feature can be looked for
in  the collective hydrodynamic
flow patterns produced when the plasma expands.

Another distinctive feature of the QGP phase diagram is shown in the right panel in 
Fig.(\ref{qgpfig2}). Recent lattice QCD calculations \cite{Fodor:2004nz}
 have begun to converge on
numerical evidence that the QGP may have a second tricritical point
\cite{Halasz:1998qr,Rischke:2003mt} 
at moderate baryon densities with $\mu_B=3\mu_q\sim 360$~MeV
and $T\sim T_c$.  

QCD predictions of the QGP phase date back thirty years \cite{Collins:1974ky}
and followed immediately after the discovery of asymptotic freedom of QCD. 
The experimental strategies to search for new forms of dense matter 
also date back thirty years when T.D. Lee proposed ``vacuum engineering'' \cite{Lee:ma}. 
It was then also 
realized by W. Greiner and collaborators \cite{Hofmann:by,Stocker:bi,Stocker:ci} 
that extended regions of dense nuclear matter 
can be formed in high energy interactions of heavy nuclei,
and that the measurement of collective flow patterns will provide the novel
barometric probes of the equation of state of ultra-dense matter. 
The hunt for the QGP and other phases of nuclear matter has been underway since
that time using several generations of higher energy accelerators,
BEVALAC, AGS, SPS, and now RHIC, and covering an impressive energy range
$\sqrt{s}-2m_N=0.2-200$~AGeV. In three years, LHC is expected to start vacuum engineering
at $5500$~AGeV.

The first conclusive evidence for (highly dissipative) 
collective nuclear flow was seen
at the BEVALAC in 1984 \cite{Reisdorf:1997fx}, and at AGS and SPS there after.
However, the first conclusive evidence for nearly dissipation free 
collective flow obeying $P_{QCD}$ had to await RHIC. In these
lectures, the discovery of the novel low dissipation
elliptic flow pattern at RHIC is highlighted as the first of three 
lines of evidence for QGP production at RHIC.
Together with two other convergent lines of evidence, 
jet quenching and the critical $D+Au$ null control, 
I conclude that the QGP has not only been discovered but
that a few of its remarkable properties have already been established
experimentally.

\section{The Empirical QGP}

The discovery of the gedanken  QGP phase of matter in the laboratory 
requires an empirical definition of the minimal number
of necessary and sufficient conditions in terms of experimentally accessible
 observables. My empirical definition is summarized by the following symbolic equation
\begin{equation}
{\rm \bf QGP= P_{QCD} + pQCD + dA}
\;\; .
\label{qgpdef}
\end{equation}
Why are three independent lines of evidence needed? 
The first term, $\underline{\rm \bf P_{QCD}}$, stands for a class of observables
that provide information about its bulk thermodynamic equation of state.
The equation of state characterizes its  {\em long wavelength}
 nonperturbative thermodynamic properties  briefly reviewed 
in section~1. 

The second term,  \underline{\rm \bf pQCD},
stands for  class of observables that provide direct evidence about its 
{\em short wavelength} dynamics predicted by  perturbative QCD. 
The QCD plasma  differs qualitative 
from familiar abelian QED plasmas due to its unique 
 non-Abelian color field dynamics. The radiative energy loss 
of energetic short wavelength 
partons was predicted to lead to striking quenching patterns 
\cite{Gyulassy:2003mc}-\cite{Wang:1992xy} of moderate and 
high  $p_T$ hadrons. 
The high RHIC cm energy of  $200$ AGeV insures that $p_T\sim 10-20$ GeV 
jet production rates are large enough to measure
via a wide array  of 
inclusive and correlation observables.
These hard partons serve as effective ``external'' tomographic probes of the  
the QGP and test its \underline{\rm \bf pQCD} {\it chromo}-dynamics.
Jets play the analogous role of neutrinos that probe the physics of stellar cores,
while hadrons play the role of photons that probe the corona of the fireball.

Below RHIC energies, the {\em \bf pQCD} line of evidence
could not be  fully developed 
because the jet rates decrease too rapidly with energy.
However, even more importantly, at the lower $p_T < 4$ GeV available
the effects of initial state nuclear dynamics and the final state
hadronic dynamics could not be 
completely  
deconvoluted from
the final spectra. This is the key point that I will repeatedly 
emphasize which differentiates the observables at the SPS and RHIC energies.
The necessity to test for the same complications at RHIC
is what gives rise to the third term in Eq.(\ref{qgpdef}).

The third term, denoted by $\underline{\rm \bf dA}$, stands for 
control experiments that can clearly differentiate
between alternative nuclear dependences specific to 
 {\it initial state} partonic wavefunctions
as well as the production mechanisms. 
The control differential, ${\rm \bf dA}$, is critical at any energy because 
the QCD plasma must first be created from pure kinetic energy! 
There has been no hot QGP in the universe (except in cosmic ray collisions)
 since the last drop condensed into
hadrons about 13 billion years ago. Cold crystalline quark matter may lurk in the cores of 
neutron stars, but the transient hot QGP 
must be ``materialized'' in the lab. The ``matter'' arises from
decoherence of {\it virtual} quantum
chromo fluctuations in the initial wavefuntions of high energy nuclei.

At ultra-relativistic energies, these virtual fluctuations are frozen out due to time dilation
into what has been  called a Color Glass Condensate (CGC) \cite{McLerran:2004fg}.
The CGC is the high density generalization of the Bjorken-Feynman dilute parton model.
At high field strengths, the non-linear interactions of virtual quantum color fluctuations
are predicted to limit the very small Bjorken $x_{BJ}\rightarrow 0$ Fourier components.
The saturation property of CGC is related to unitarity constraints and 
determines the maximal entropy that can be produced in $AA$ at a given $\sqrt{s}$
as also pointed out by EKRT~\cite{Eskola:2002qz}. 
 
The ${\rm \bf dA}$ control is needed to characterize to what extent these nonlinear initial
state physics effects can be differentiated from effects due to final state interactions
in the QGP matter that forms from it.
At RHIC, the best experimental handle on the ${\rm \bf dA}$ term 
happens to be the study of $D+A$ reactions. In such light-heavy ion reactions,
the initial state CGC physics can be isolated 
because the produced QGP, if any, is too tenuous.

Why don't I add more terms in Eq.(\ref{qgpdef})? In fact, each term stands for
many independent components, as I elaborate below.
For the three required terms in Eq.(1) the published experimental
evidence is now overwhelming and conclusive. 
Four independent experiments have converged to complementary 
very high quality data sets. 

The three terms in Eq.(1) are  necessary and sufficient for 
establishing that a  discovery has been made of a uniquely different
form of strongly interacting QGP.
After discussing the three lines of 
evidence, I will elaborate on why I believe that 
direct photons, $J/\psi$, HBT, or other interesting 
observables do not need to be added to Eq.(1). Those observable provide
valuable additional constraints on the {\em combined and convoluted} 
properties of the initial state, the  QGP, {\em AND} 
the dense hadronic matter into which it condenses.
However, the deconvolution of the initial and hadronic final
effects has already proven to be very difficult at SPS energies
and will continue to be at RHIC. 

To avoid misunderstanding, the discovery of the QGP does not mean
that its physical properties are now understood. In fact, it only signals 
the beginning of a long and well focused direction of research. 
The history of the neutron star discovery offers an instructive analogy.
In 1934 Baade and Zwicky proposed the theoretical existence of neutron stars
soon after Chadwick discovered free neutrons. Thirty years later 
in 1967 Hewish and Bell observed the first few pulsars when suitable radio telescopes
could finally be constructed. An amusing anecdote is that
they actually agonized for a time about whether LGM 
(little green men) were sending them encrypted
 messages from the cosmos. T.~Gold in 1968 
(as $D+Au$ did at RHIC in 2003) put the debate to rest. Gold proposed that
 radiative energy loss of a magnetized neutron star 
would cause a predictable spin down. Later 
precision measurements confirmed this. Seventy years after its proposal, neutron star
research still remains a very active experimental and theoretical direction of physics.
Current interest has focused on 
possible {\em color field} super-conductivity~\cite{Alford:1999pb} 
recently predicted in the very high $\mu_B$ sector of the QCD phase diagram,
beyond the boundaries~\cite{Rischke:2003mt} of Fig.2b.

The critical $D+Au$ control experiments in 2003 could have 
found that the $Au+Au$ QGP observables  were strongly distorted by the 
possible initial CGC 
state that created it. This would have certainly foiled Eq.(1).
The search for the bulk  
QGP phase of matter would then have had to await higher energies and densities at LHC
or for a better understanding of how to deconvolute that initial state 
physics. The large positive signatures in similar $p+Pb$ control experiments at SPS 
showed in fact initial effects strongly distort key observables.
At SPS the physics of high $p_T$ Cronin enhancement
and $p+A\rightarrow J/\psi$ suppression remain the important open problems.
In contrast, at RHIC energies, the absence of jet quenching at midrapidity 
and the ``return of the jeti'' correlations in ${\bf dA=D+Au}$
provided the check-mate completion of Eq.(1).

As emphasized by McLerran\cite{McLerran:2004fg} in these proceedings,
the $D+Au$ control at RHIC at high rapidities does in fact produce a positive
signature for new  initial state  physics.  In those
kinematic ranges  $x_{BJ}<0.001$,  ${\rm  {\bf dA}}$ fails as a null control for QGP,
but may signal the onset of nonlinear CGC initial conditions.
In this lecture, I concentrate on the midrapidity region, $x_{BJ}>0.01$, where
Eq.(1) was conclusively satisfied.

\section{${\rm \bf P_{QCD}}$ and Bulk Collective Flow}

The identification of a new form of ``bulk matter''  requires the 
 observation of novel and uniquely different
collective properties from ones seen before.
This  requirement is the first term in Eq.(1).  
In heavy ion collisions, the primary observables of bulk
collectivity are the radial, azimuthal
and longitudinal flow patterns
of hundreds or now thousands of produced hadrons.
Stocker, Greiner,  and collaborators were the first to predict
\cite{Hofmann:by,Stocker:bi,Stocker:ci,Stocker:vf,Stocker:pg}
distinctive ``side splash and squeeze-out'' collective 
flow patterns in nuclear collisions.
The different types of collective 
flows are conveniently quantified in terms of the first 
few azimuthal Fourier components~\cite{Ollitrault:bk}, $v_n(y,p_T,N_p,h)$,
 of centrality selected triple differential inclusive distribution
of hadrons, $h$. The centrality or impact parameter range is usually
specified by a range of associated multiplicities, from which
the average number of participating nucleons, $N_p$, can be deduced.
The azimuthal angle of the hadrons are measured relative to 
 a globally determined estimate for the collision reaction plane angle 
$\Phi(M)$. The ``directed'' $v_1$ and ``elliptic'' $v_2$ flow components
\cite{Reisdorf:1997fx,Ollitrault:bk,Voloshin:1999gs}-\cite{Back:2002ft}  are
readily identified from azimuthal dependence
\begin{eqnarray}
\frac{dN_h(N_p)}{dydp_T^2d\phi}
=\frac{dN_h(N_p)}{dydp_T^2} \frac{1}{2\pi}(1 &+& 2 v_1(y,p_T,N_p,h) 
\cos\phi  \nonumber
\\   &+&  2 v_2(y,p_T,N_p,h) \cos 2 \phi + \cdots ) \;\;.
\label{floweq}
\end{eqnarray}
The ``radial flow'' component, ``1'' , is  identified~\cite{Cheng:2003as}
from the hadron mass dependence of the  blue shifted transverse 
momentum spectra
\begin{equation}
\frac{dN_h(N_p)}{dydp_T^2}\sim \exp[ -m_h\cosh(\rho_\perp-\beta(y))/T_f] \;\;,  
\end{equation}
where $m_h(\sinh(\rho_\perp),\cosh(\rho_\perp))=(p_\perp,\sqrt{m_h^2+p_\perp^2})$
and $\beta(y)$ is the mean collective transverse flow rapidity at $y$.
\begin{figure}[h]
\centering
\includegraphics[height=0.35\textheight,width=0.47\textwidth,clip]{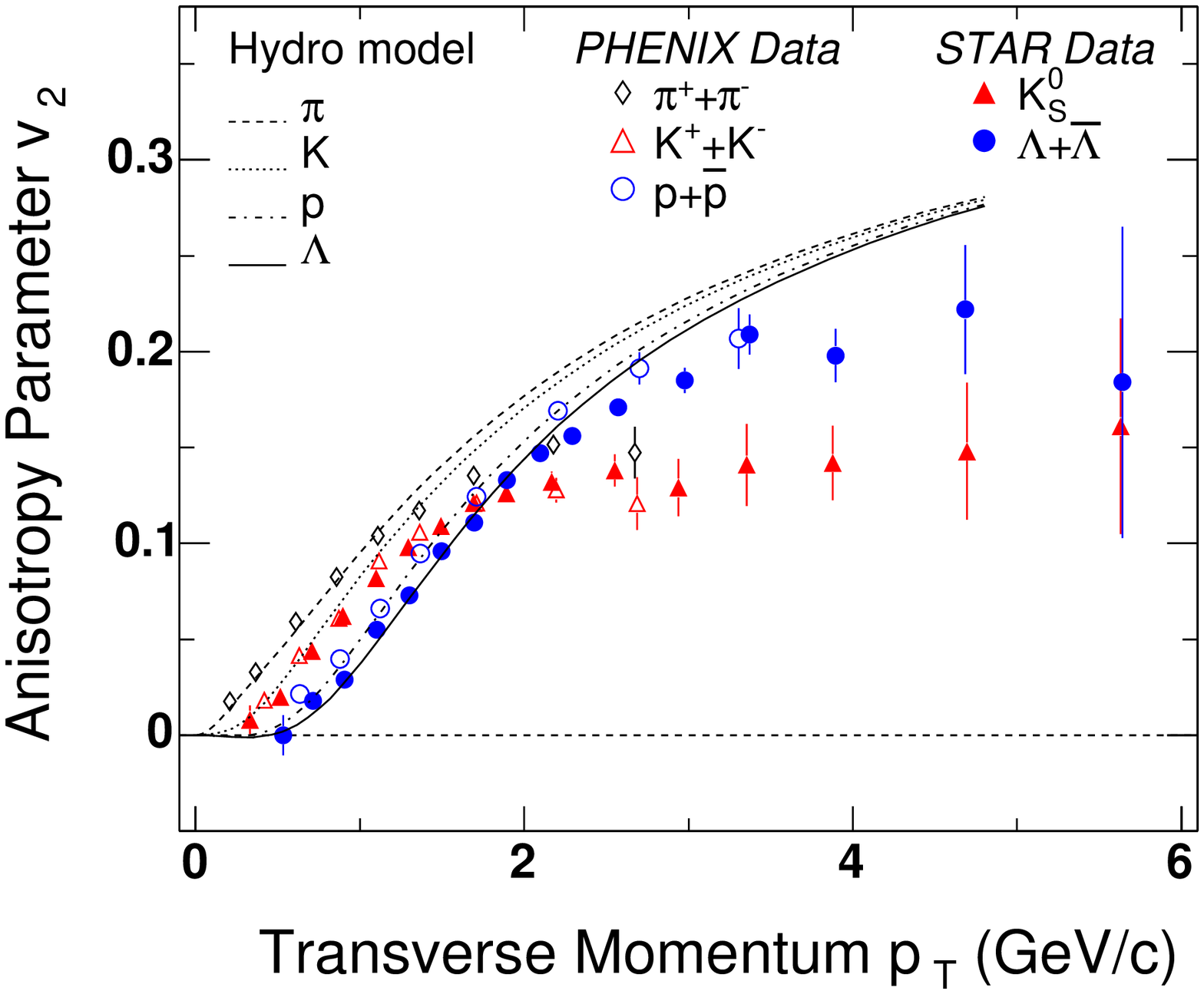}
\hfill
\includegraphics[height=0.35\textheight,width=0.5\textwidth,clip]{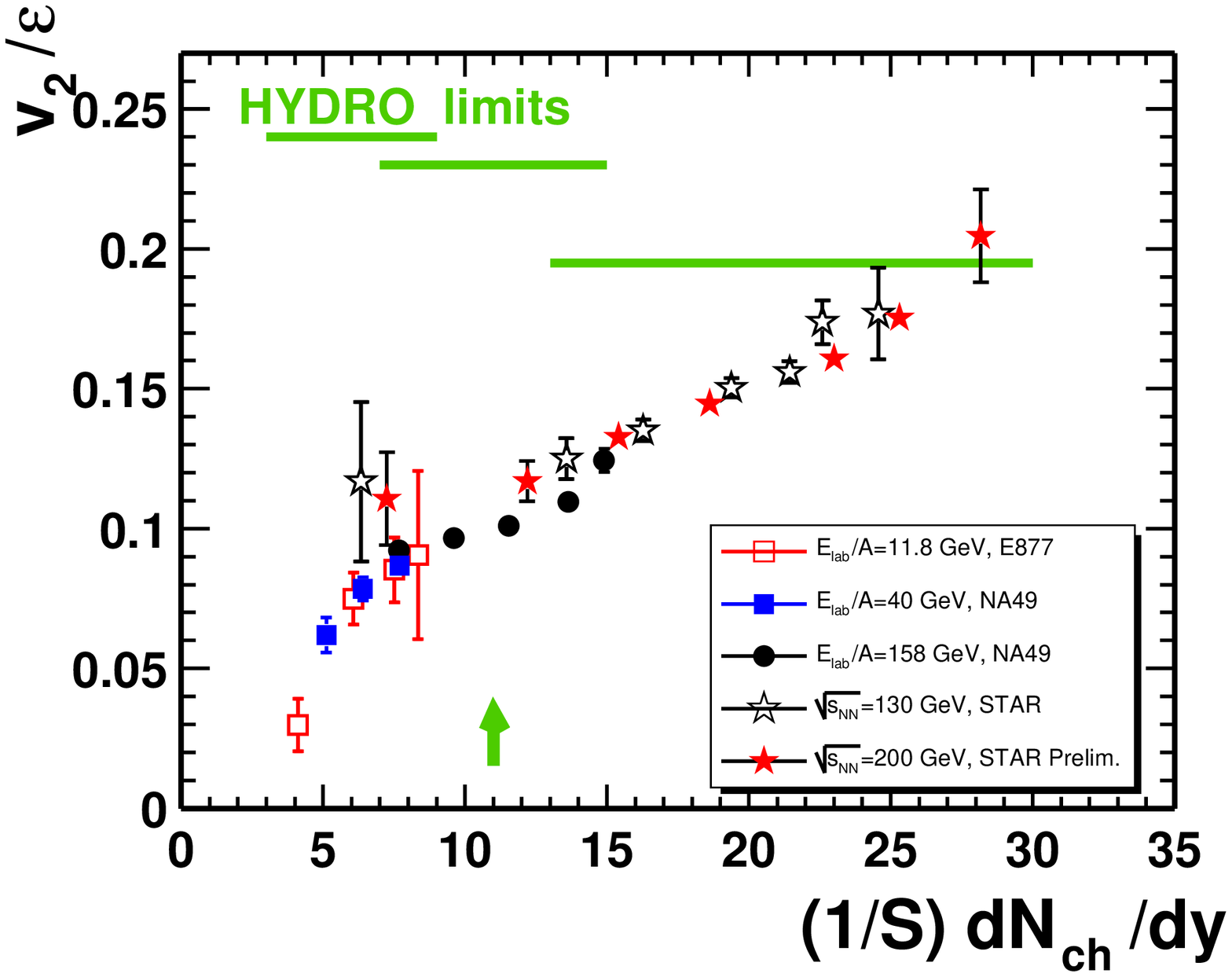}
\caption{First line of evidence: Bulk collective flow is the barometric signature
of QGP production. Left figure combines STAR~\protect{\cite{Adams:2003zg}-\cite{Adler:2002pu}}
and PHENIX~\protect{\cite{Adler:2003kt,Adler:2003qi}} measurements of the azimuthal 
elliptic flow ($v_2(p_T)$) of $\pi,K,p,\Lambda$ in Au+Au at 200 AGeV. 
The predicted hydrodynamic flow pattern 
from \protect{\cite{Kolb:2000fh}-\cite{Kolb:2003dz}} agrees well with observations
in the bulk $p_T<1$ GeV domain. Right figure from \protect{\cite{Alt:2003ab}}
shows $v_2$ scaled to the
initial elliptic spatial anisotropy, $\epsilon$, as a function
of the charge particle density per unit transverse area. The bulk hydrodynamic limit
is only attained at RHIC.}
\label{line1fig}
\end{figure}

Figure (\ref{line1fig}) shows the striking bulk collectivity 
elliptic flow signature of QGP formation at RHIC. 
Unlike at SPS and lower energies, the observed large
elliptic deformation ($(1+2 v_2)/(1-2 v_2)\sim 1.5$)
of the final transverse momentum distribution 
agrees for the first time with 
non-viscous hydrodynamic predictions~\cite{Kolb:2000fh}-\cite{Hirano:2003pw}
at least up to about $p_T\sim 1$ GeV/c.
However, the right panel shows that when the local rapidity density per unit 
area~\cite{Voloshin:1999gs,Alt:2003ab} drops 
below the values achieved at RHIC $\sim 30/{\rm fm}^2$, 
then the elliptic flow (scaled by the initial 
spatial ellipticity, $\epsilon=\langle (y^2-x^2)/(y^2+x^2)\rangle$) falls below the 
perfect fluid hydrodynamic predictions. We will discuss in more detail the origin of
the large discrepancy at SPS energies in the next section. 

The most impressive feature in Fig.(\ref{line1fig})
is the agreement of the observed hadron mass dependence 
of the elliptic flow pattern for 
all hadron species, $\pi, K, p,\Lambda$, with the hydrodynamic predictions
 below 1 GeV/c. This is the QGP fingerprint that shows that there is a common 
 bulk collective azimuthally asymmetric flow velocity field, $u^\mu(\tau,r,\phi)$.

The flow velocity and temperature fields of a perfect (non-viscous)
fluid obeys the  hydrodynamic equations:
\begin{equation}
\partial_\mu\left\{[\epsilon_{QCD}(T(x))+P_{QCD}(T(x))]u^\mu(x)u^\nu(x)-g^{\mu\nu} 
P_{QCD}(T(x))\right\} = 0 \; ,
\label{hydro}
\end{equation}
where  $T(x)$ is the local temperature field, $P_{QCD}(T)$ is the QGP equation of state,
and $\epsilon_{QCD}(T)=(TdP/dT -P)_{QCD}$ is the local proper energy density.
The above equations apply in the rapidity window $|y|<1$,
where the baryon chemical potential 
can be neglected. 
Eq.(\ref{hydro}) provides the barometric
connection between the observed flow velocity and the sought after ${\rm \bf P_{QCD}}$.
\begin{figure}[h]
\centering
\includegraphics[height=0.3\textheight,width=0.45\textwidth,clip]{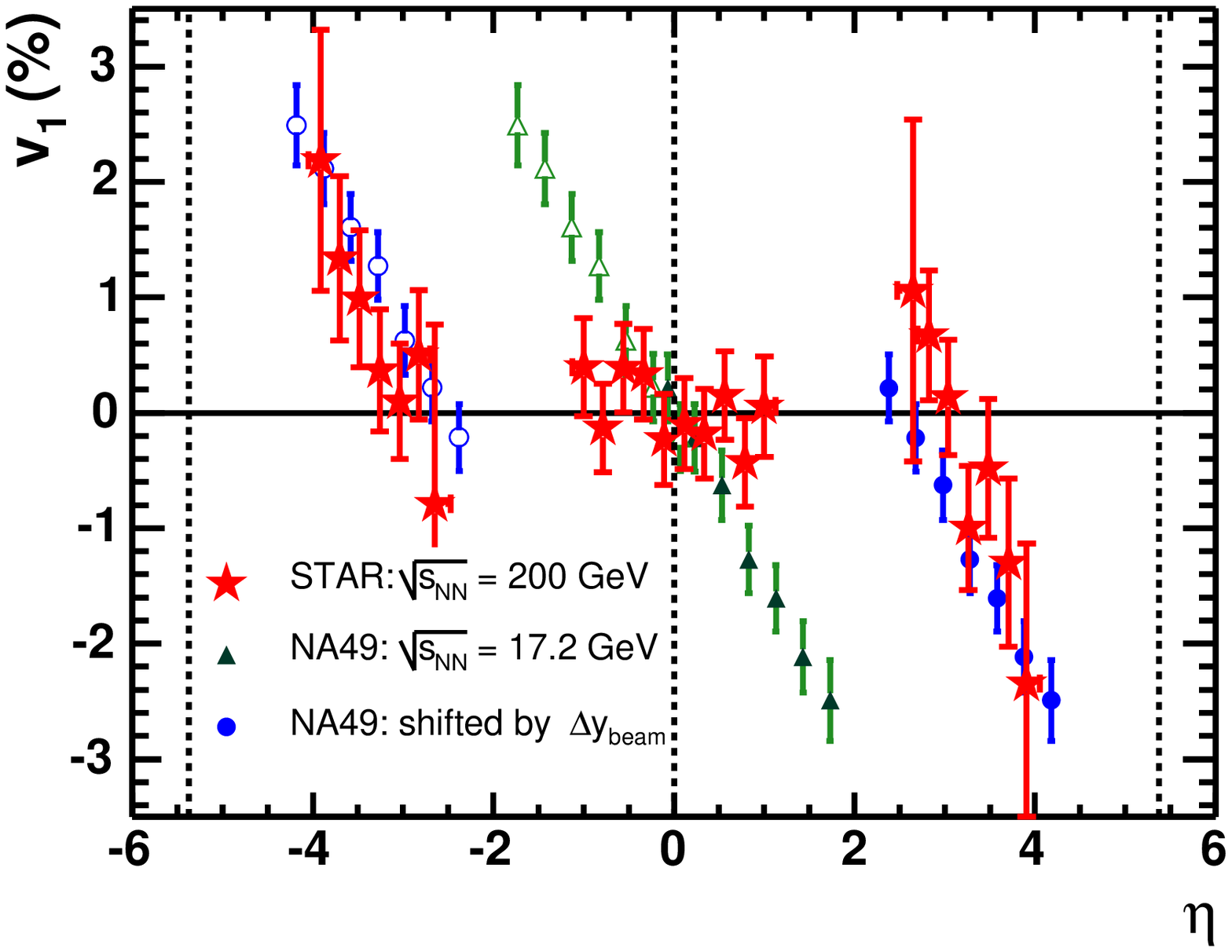}
\includegraphics[height=0.3\textheight,width=0.45\textwidth,clip]{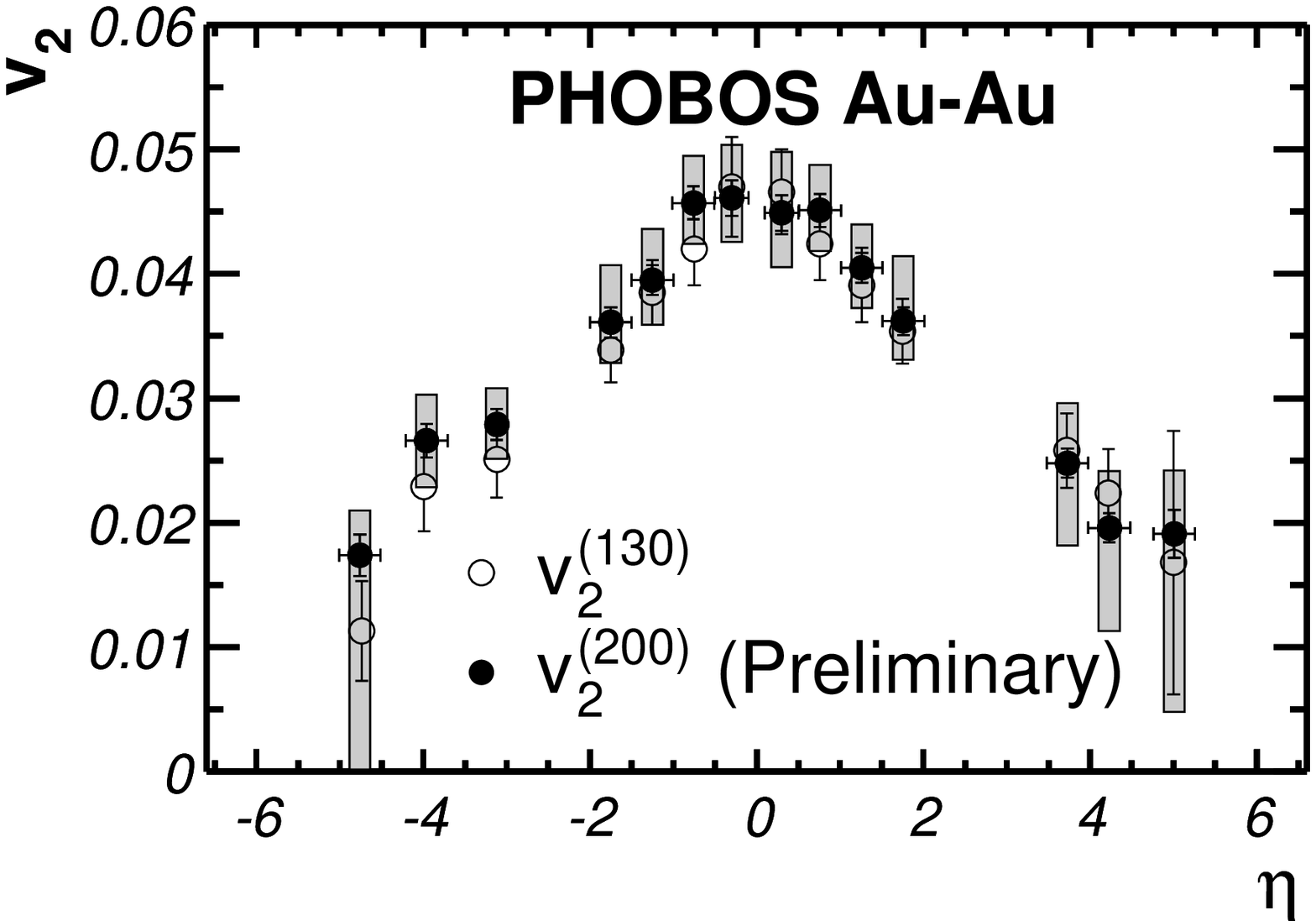}
\caption{ Left figure combines STAR and NA49 data~\protect{\cite{Adams:2003zg}}
and shows that the directed sidewards flow, $v_1(y)$, 
is correlated over 8 units of rapidity at RHIC. 
At SPS collectivity is dominated by the overlapping fragmentation regions
while at RHIC the nearly identical directed flow of in the fragmentation regions 
is shifted to $|y|>2$. Right figure shows the pseudo rapidity dependence
of elliptic from PHOBOS~\protect{\cite{Back:2002ft}}.}
\label{line1figb}
\end{figure}

The long range nature of collective flow has also been conclusively
established by STAR and PHOBOS seen in Fig.(\ref{line1figb}). 
The sidewards flow is anti-correlated over 8 units of rapidity!  In addition, its azimuthal
orientation was shown to coincide with the azimuthal direction of the largest
axis of elliptic deformation at $y=0$. This provides an important
test of the overall consistency of the hydrodynamic origin of
flow. The rapidity dependence of the elliptic flow in
Fig.(\ref{line1figb}) also shows the long range nature of bulk collectivity.

Why is $v_2$ more emphasized than $v_1$ or radial flow as a signature of QGP formation?
The primary reason is that elliptic flow is generated mainly during the 
highest density phase of the evolution before 
the initial geometric spatial asymmetry of the plasma
disappears. It comes from the azimuthal dependence of the pressure gradients,
which can be studied by varying the centrality of the events~\cite{Ollitrault:bk}. 
Detailed parton transport~\cite{Molnar:2001ux} and hydrodynamic~\cite{Teaney:2001av}
calculations show that most of the 
$v_2$ at RHIC is produced before 3 fm/c and that elliptic flow is 
relatively insensitive to the
late stage dissipative expansion of the hadronic phase.
In contrast, radial flow has been observed at all energies~\cite{Cheng:2003as}
 and has been shown to be very sensitive to late time 
``pion wind'' radial pressure gradients~\cite{Bass:2000ib}, which continue
to blow after the QGP  condenses into hadronic resonances.

The observation of  near ideal
$v_2$ fluid collectivity as predicted with the $P_{QCD}$ together with $v_1(y)$ and 
other consistency checks (${\bf \rm\bf c.c.}$) 
conclusively establish the first term in Eq.(1) :
\begin{equation}
{\rm\bf P_{QCD}}= v_2(p_T; \pi,K,p,\Lambda) + v_1(y) + \;{ \rm\bf c.c.}
\; \; .
\label{Pdef2}
\end{equation}
Preliminary Quark Matter 2004  analysis of $\Xi,\Omega$ flow are consistent
with the predicted $v_2(p_T,y=0,M,h)$ and this information 
is lumped into the ${\bf \rm\bf c.c.}$ terms
in Eq.(\ref{Pdef2}). Other data which provide consistency checks
of the hydrodynamic
explanation of collective flow include the 
observed $\pi,K,p$ radial flow data~\cite{Cheng:2003as} for $p_T<2$ GeV.
In addition, predicted statistical thermodynamic distributions~\cite{Braun-Munzinger:2003zd} 
of final hadron yields agree remarkably well with RHIC data.
Had hadro-chemistry failed at RHIC, then a large question mark would have
remained about bulk equilibration in the QGP phase.

\section{QGP Precursors at SPS and Dissipative Collectivity}

It is important to point out, 
that no detailed 3+1D hydrodynamic calculation~\cite{Hirano:2003hq}-\cite{Hirano:2003pw}
has yet been able to reproduce the 
rapid decrease of $v_2(|\eta|>1)$ observed by  PHOBOS in Fig.(\ref{line1figb}). 
This discrepancy is due, in my opinion, to the onset of  {\em hadronic} dissipation 
effects as the comoving density
decreases with increasing $y$. 
From the right panel of Fig.(\ref{line1fig}), we see that as a decrease of the local
transverse density from midrapidity RHIC conditions 
leads to an increasing deviation from the perfect fluid limit.
The initial density was also observed to decrease 
at RHIC as $|y|$ increases~\cite{Bearden:2001qq}.
Therefore, from SPS data alone, we should have expect deviations from the 
perfect fluid limit away from the midrapidity region.
It would be interesting to superpose the PHOBOS data 
on top of the NA49 systematics.

\begin{figure}[h]
\centerline{\includegraphics[width=5cm,height=6cm]{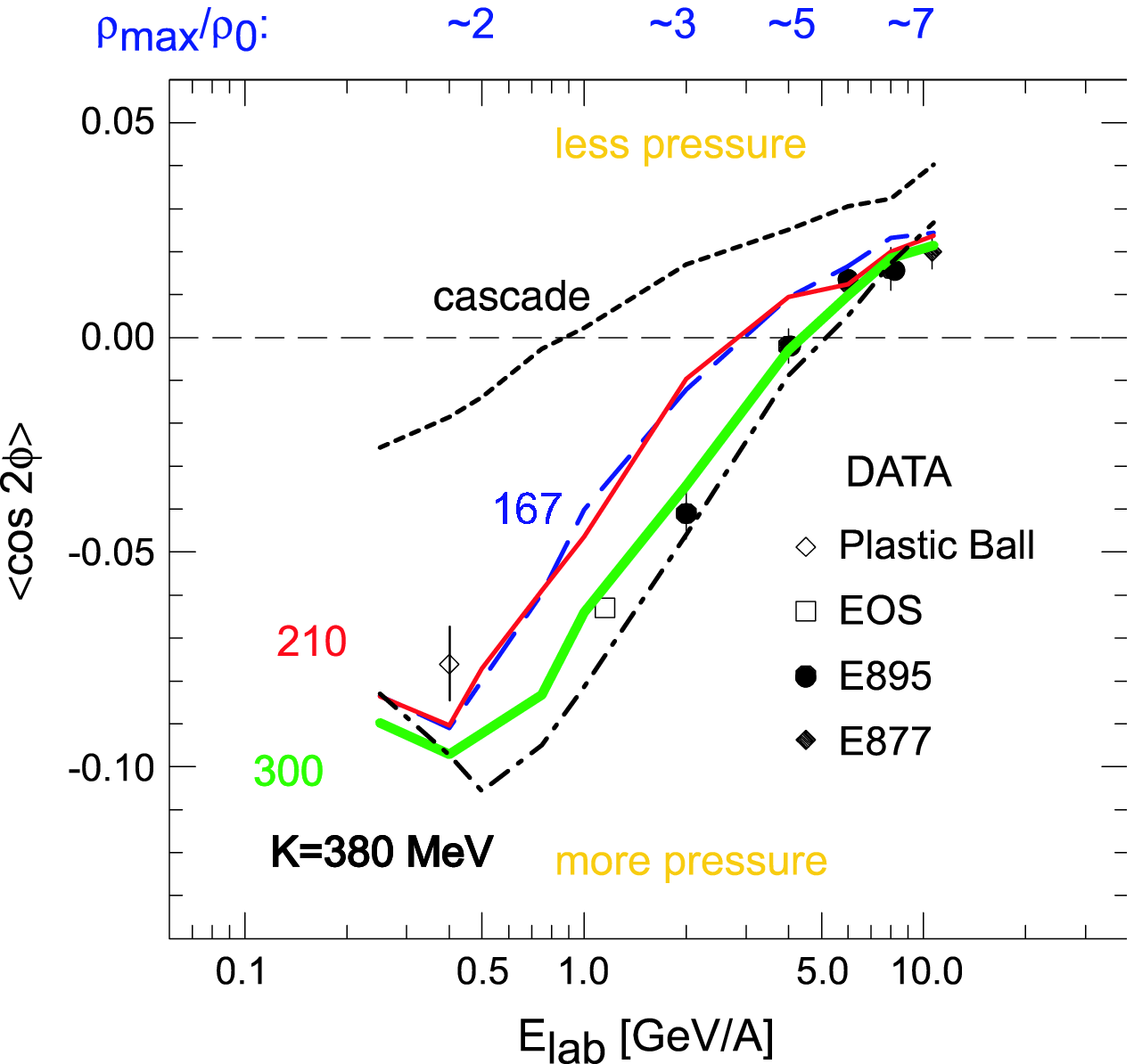}
\includegraphics[width=5cm]{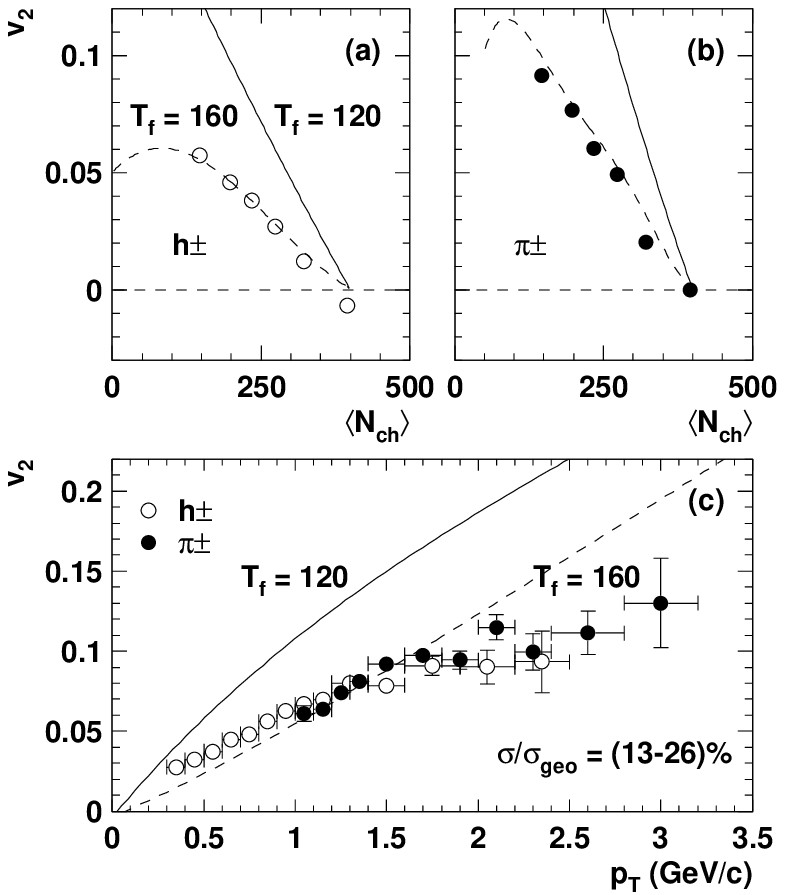}}
\caption{Evidence for dissipative collective flow below RHIC energies.
Left: Non equilibrium BUU nuclear transport theory {\protect
\cite{Danielewicz:2002pu,Stoicea:2004kp}} can explain the observed
elliptic squeeze-out (negative $v_2$) collectivity below 4 AGeV. 
Right: CERES{\protect\cite{Agakichiev:2003gg}} 
data on elliptic flow at SPS is well below hydrodynamic predictions 
with freeze-out $T_f=120$ MeV required to reproduce the single inclusive
radial flow. Early freeze-out with $T_f=160$ MeV, simulating effects of
dissipation, is needed to reproduce the data.}
\label{spsv2}
\end{figure}

To elaborate on this point, Fig.\ref{spsv2} shows CERES data~\cite{Agakichiev:2003gg} on
 $v_2(p_T)$  at SPS energy $\sqrt{s}=17$ AGeV. In agreement with the right panel
of Fig.(\ref{line1fig}), the CERES data falls well below hydrodynamic
predictions. At even lower energies, AGS and BEVALAC,
the $v_2$ even  becomes {\em negative} and this ``squeeze out'' of plane~\cite{Stocker:ci}
 is now 
well understood in terms of non-equilibrium BUU 
nuclear transport theory~\cite{Stoicea:2004kp,Danielewicz:2002pu}.
In order to account for the smallness of $v_2$ at SPS, hydrodynamics has to be frozen
out at unphysically high $T_f\approx T_c=160$ MeV. However,
the observed radial flow rules out
this simple fix. 

The discrepancy of $v_2$ and hydro at SPS energies
can be traced to the important contribution
of the dissipative final {\em hadronic}
state interactions. The hadronic fluid is far from ideal.
In approaches~\cite{Bass:2000ib, Teaney:2001av} 
that combine perfect fluid
QGP hydrodynamics with non-equilibrium hadronic transport
dynamics, the  
importance of dissipative hadron dynamics at SPS was clearly demonstrated.
The problem is that the QGP at lower initial densities 
condenses on a faster time scale cannot take advantage of the 
of the  spatial asymmetry to generate large $v_2$.
The subsequent dissipative hadronic fluid is very inefficient in exploiting
spatial asymmetry.
A factor of two reduction of the initial QGP density, therefore,
 leads to a significant 
systematic bias of the $v_2$ barometer, not only at SPS but also at high $|y|$
at RHIC. Current 
hadronic transport theory is not yet accurate enough to re-calibrate the  barometer
away from mid-rapdities at RHIC.

In light of the above discussion, the smallness of dissipative corrections in the central
regions of RHIC is even  more  surprising.
At mid-rapidities, the lack of substantial dissipation in the QGP phase 
is in itself  remarkable.
Calculations based on parton transport
theory~\cite{Molnar:2001ux} predict large deviations from the ideal
non-viscous hydrodynamic limit.  Instead, the data show that
the QGP is almost a perfect fluid. A Navier Stokes analysis~\cite{Teaney:2003pb}
is consistent with \cite{Molnar:2001ux} and shows that the 
viscosity of the QGP must be about ten times less than 
expected if the QGP were a weakly interactive Debye screened plasma.
This unexpected feature of the QGP must be due to nonperturbative 
and hence strong coupling physics that persists to at least $3T_c$.

\subsection{The Minimal Viscosity of the QGP}

One intriguing theoretical possibility being explored in the literature
\cite{Policastro:2002tn}
is that the shear viscosity, $\eta$, in the strongly coupled QGP may saturate 
at a universal
super-string bound, $\eta/\sigma=1/4\pi$. This conjectured
duality between string theory and QCD may help to explain also
the $\sim 20\%$ deviation of $P_{QCD}(T)$ from the ideal Stefan Boltzmann limit.
The discovery of nearly perfect fluid flow of {\em long wavelength} 
modes with $p_T<1$ GeV at RHIC is certain to fuel more
interest in this direction.

I propose that a simpler physical explanation of the 
lower bound on viscosity follows from the
uncertainty principle, as derived in Eq.(3.3) of Ref.~\cite{Danielewicz:ww}. 
Standard kinetic theory derivation of shear viscosity leads to
$\eta=(\rho\langle p\rangle \lambda)/3$ where $\rho$ is the proper density,
$\langle p \rangle$ is the average total momentum, and $\lambda$ is the momentum degradation
transport mean free path. The uncertainty principle implies that quanta with average momentum
components  $\langle p \rangle$ cannot be localized to better than $\Delta x\sim
1/\langle p \rangle$. Therefore the momentum degradation mean free path 
cannot be defined more accurately than
$ \lambda>1/\langle p \rangle$. For an ultra-relativistic system, the 
entropy density is $\sigma\approx 4 \rho$, therefore
\begin{equation}
\frac{\eta}{\sigma}\stackrel{~}{>} \frac{1}{12} \;\;, 
\label{etabound}
\end{equation}
which  is within 5\% of the string theory bound.
It is the consequence of the universality 
of the Heisenberg uncertainty principle. 
Surprisingly, the QGP found at RHIC saturates this uncertainty bound
and the data clearly rule out the order of magnitude 
larger predictions based on pQCD~\cite{Danielewicz:ww}. See 
again~\cite{Molnar:2001ux}.  The long wavelength
modes in the QGP are as maximally coupled as $\hbar=1$ allows.

It is ``shear'' good luck that the mid rapidity initial conditions at RHIC are dense
enough to essentially eliminate the dilution of elliptic flow
due to the imperfect hadron fluid formed after the spatial asymmetry vanishes.
(Recall that $(\eta/\sigma)_H\sim (T_c/T)^{1/{c_H}^2}>1$ for $T<T_c$  \cite{Danielewicz:ww}).
At lower energies or higher rapidities this good luck runs out
and the mixture of near perfect QGP and imperfect hadronic fluid dynamics
reduces the elliptic flow.

\section{ pQCD and Jet Quenching}

In addition to the breakdown of perfect fluid collectivity at high rapidity
seen in Fig.(\ref{line1figb}), 
Fig.(\ref{line1fig}) clearly shows 
that hydrodynamics also breaks down at  very short wavelengths and
high transverse momenta, $p_T> 2$ GeV. Instead of continuing to rise with $p_T$,
the elliptic asymmetry stops growing and the difference between baryon vs meson 
$v_2$ reverses sign! Between $2<p_T<5$ GeV the baryon $v_2^B(p_T)$ exceeds the
meson $v_2^M(p_T)$ by approximately 3/2. 
For such short wavelength components of the QGP, local equilibrium simply cannot be
maintained due the fundamental asymptotic freedom property
of QCD.  I return to the 
baryon dominated transition region $1<p_T<5$ GeV
in a later section since this involves interesting but as yet  uncertain non-equilibrium
non-perturbative processes.
In this section I concentrate on the $p_T>2$ GeV meson observables 
that can be readily understood in terms QGP modified
{\bf pQCD} dynamics~\cite{Gyulassy:2003mc,Baier:2000mf}. 

The quantitative study of short wavelength partonic {\bf pQCD} dynamics 
focuses on the rare high $p_T$
power law tails that extend far beyond the typical (long wavelength) 
scales $p<  3 T \sim 1$ GeV
of the bulk QGP. The second major discovery at RHIC
is that the non-equilibrium power law high $p_T$ jet distributions remain power law like 
but are strongly quenched~\cite{Adcox:2001jp}-\cite{Adler:2002ct}.
Furthermore, the quenching pattern has a distinct centrality, $p_T$, azimuthal angle,
and hadron flavor dependence that can be used to test the 
underlying dynamics in many independent ways.
\begin{figure}[htp]
\centering
\includegraphics[height=0.3\textheight,width=0.47\textwidth,clip]{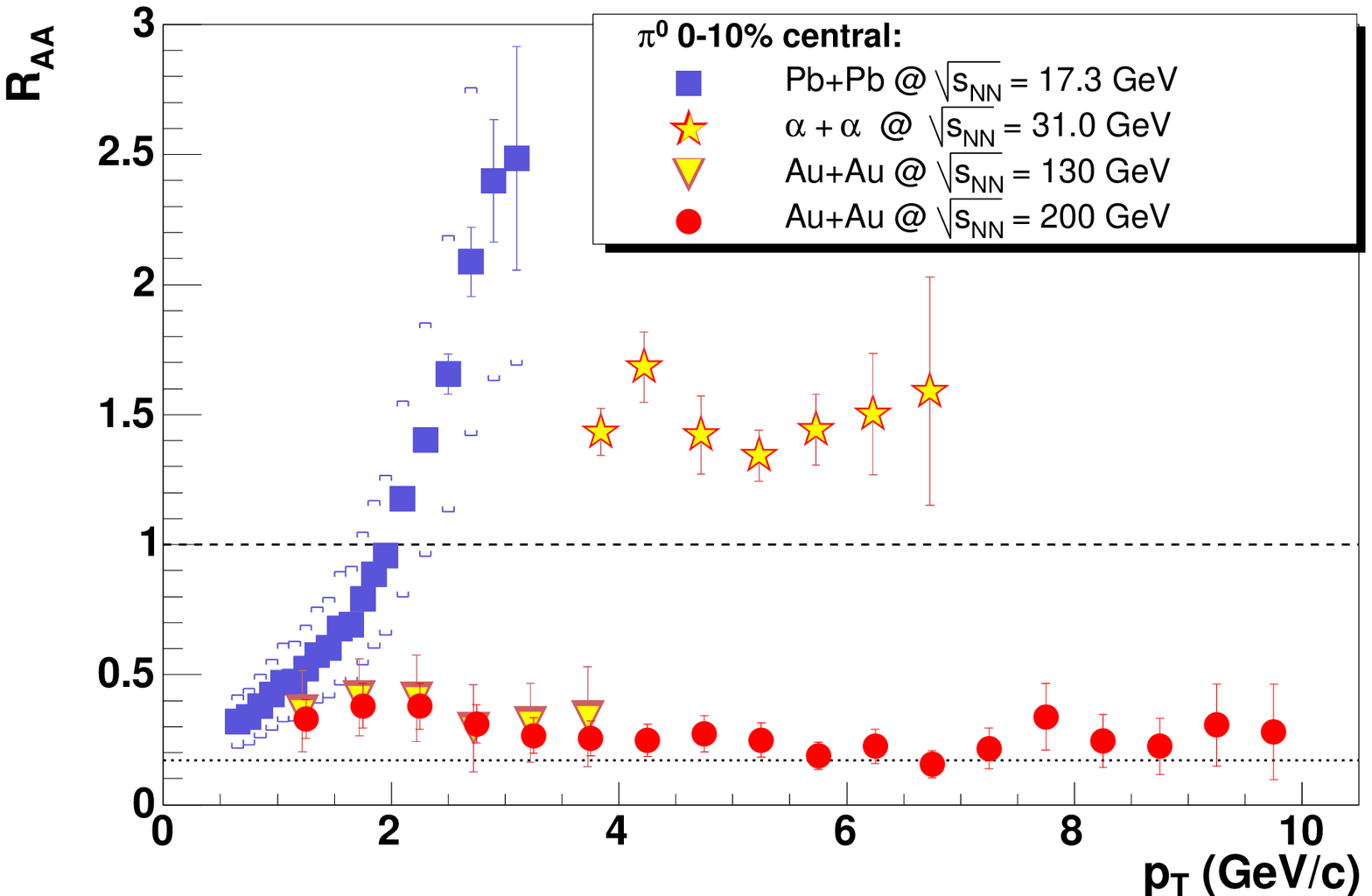}
\includegraphics[height=0.31\textheight,width=0.5\textwidth,angle=-90,origin=c,clip] {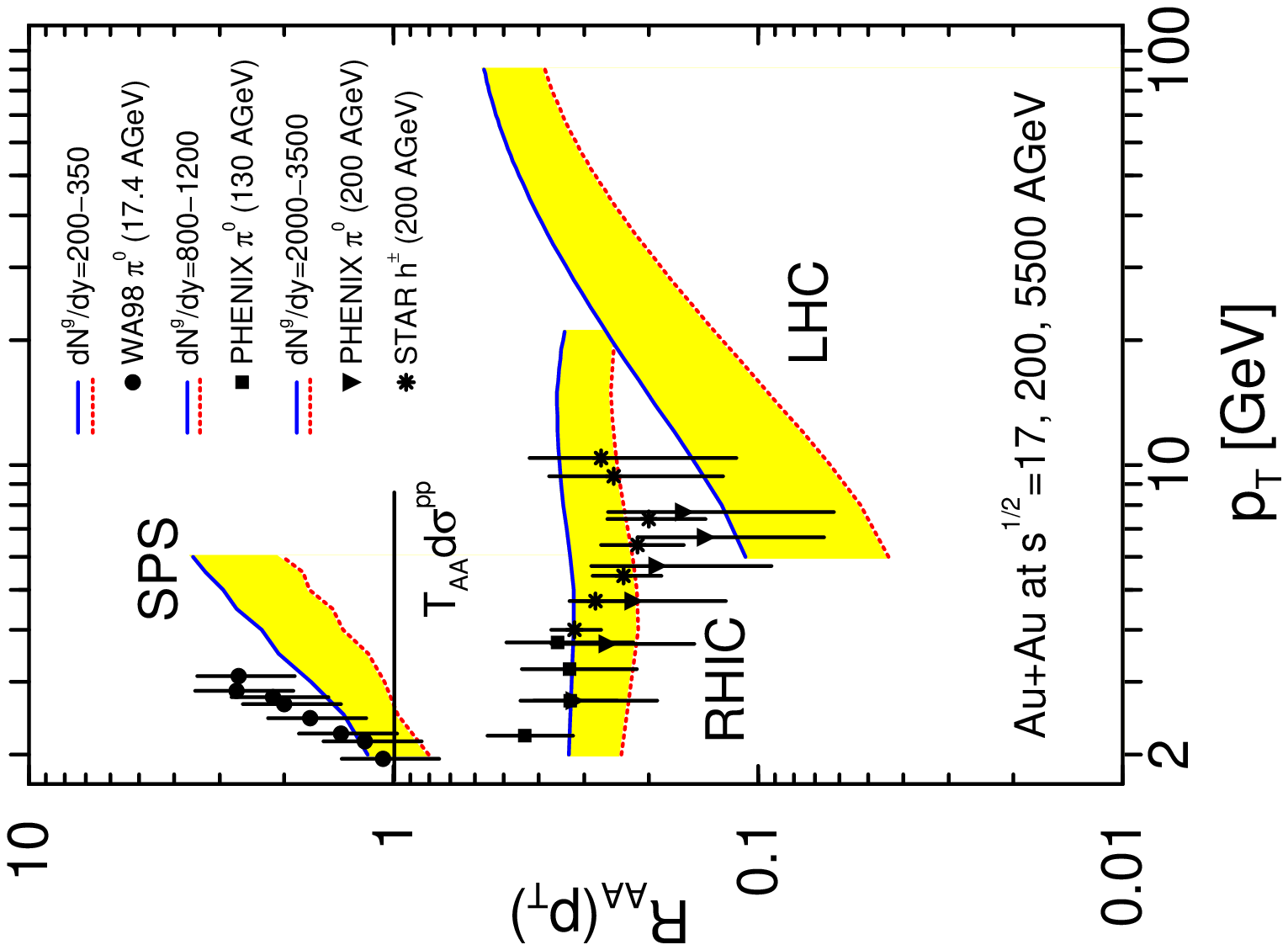}
\caption{Jet Quenching at RHIC.
Left \protect{\cite{d'Enterria:2004ne}} shows the jet quenching pattern of $\pi^0$ 
discovered by PHENIX~\protect{\cite{Adcox:2001jp,Adcox:2002pe}}
 at RHIC compared to previous observation of high $p_T$ enhancement
at ISR and SPS energies. The nuclear modification factor
$R_{AA}= dN_{AA}/T_{AA}(b)d \sigma_{pp}$ measures the deviation of $AA$ spectra
from factorized pQCD. Right shows predictions~\protect{\cite{Vitev:2002pf}} 
of the $\sqrt{s}$
and $p_T$ dependence from SPS, RHIC, LHC based on the GLV 
theory~\protect{\cite{Gyulassy:2000er}} of radiative energy loss.}
\label{line2fig}
\end{figure}
Below RHIC energies, the initial state Cronin enhancement of moderately
high $p_T$ tails was observed in central $Pb+Pb$ reactions at the SPS.
At the ISR a reduced Cronin enhancement in $\alpha+\alpha$ reactions
was seen.  In contrast, at RHIC a 
large suppression, by a factor of 4-5, was discovered in central $Au+Au$ that extends 
beyond 10~GeV for $\pi^0$.  

Jet quenching in $A+A$ was proposed in~\cite{Gyulassy:1990bh,Wang:1992xy} 
as a way to study the dense matter produced at RHIC energies. 
As noted before, the pQCD 
jet production  rates finally become large enough to measure yields up to 
high $p_T > 10$ GeV.
Order of magnitude suppression effects were predicted based on simple
estimates of induced gluon radiative energy loss.
Ordinary, elastic energy loss~\cite{Bjorken:1982tu} 
was known by that time to be too small to lead to significant attenuation.

As reviewed in~\cite{Gyulassy:2003mc,Baier:2000mf}
refinements in the theory since then have opened the possibility of 
using the observed jet quenching pattern
as a tomographic tool~\cite{TOMO} 
to probe the parton densities in a QGP. The right panel shows
a recent jet tomographic analysis~\cite{Vitev:2002pf} of the PHENIX 
$\pi^0$ data~\cite{Adcox:2001jp,Adcox:2002pe}
based on the GLV opacity formalism~\cite{Gyulassy:2000er}.
Vitev and I concluded from  Fig.6b that the 
initial gluon rapidity density 
required to account for the observed jet quenching pattern must be
$dN_g/dy\sim 1000\pm 200$. 

This jet tomographic measure of the initial $dN_g/dy$ 
is in remarkable agreement with three other independent
sources: (1) the initial  entropy
deduced via the Bjorken formula from the measured multiplicity, (2) 
the initial condition of the QGP required in hydrodynamics 
to produce the observed elliptic flow, 
and  (3) the 
estimate of the maximum gluon rapidity density
bound from the CGC gluon saturated initial condition \cite{Eskola:2002qz}.

These four independent measures makes it possible to estimate 
the maximal initial energy density in central collisions 
\begin{equation}
\epsilon_0 = \epsilon(\tau \sim 1/p_0) \approx \frac{p_0^2}{\pi R^2}\frac{dN_g}{dy}
\approx 20 \frac{{\rm GeV}}{{\rm fm}^3} \sim 100 \times \epsilon_{A}
\label{eps0}
\end{equation}
where $p_0\approx Q_{sat}\approx 1.0-1.4$ GeV is the mean transverse momentum
of the initial produced gluons from the incident saturated virtual nuclear
CGC fields\cite{McLerran:2004fg,Eskola:2002qz}. 
This scale controls the
formation time $\hbar/p_0\approx 0.2$ fm/c of the initially out-of-equilibrium 
(mostly gluonic) QGP.  
The success of the hydrodynamics
requires that local equilibrium  be achieved on a 
fast proper time scale $\tau_{eq}\approx (1-3)/p_0 < 0.6$ fm/c.
The temperature at that time is 
$T(\tau_{eq})\approx (\epsilon_0/(1-3)\times 12)^{1/4} \approx 2 T_c$.

In HIJING model\cite{ToporPop:2002gf}, 
the mini-jet cutoff is $p_0=2-2.2$ GeV  limits the number of mini-jets
well below 1000. The inferred opacity of the QGP is much higher and consistent with 
the CGC and EKRT estimates.

\subsection{\protect{$R_{AA}(p_{T})$} and Single Jet Tomography}

In order to illustrate the ideas behind jet tomography, 
I will simplify the discussion here to a schematic form.
See \cite{Gyulassy:2003mc} for details.
The fractional  radiative energy loss of a high energy parton 
in an expanding QGP
is proportional
to the position weighed line integral over color charge
density 
$\rho(\vec{x}_\perp,\tau)$
\begin{equation}
\Delta E_{GLV}/E \approx  C_2 \kappa(E)  \int_{0}^{L(\phi)} d\tau\; 
\tau\rho(\vec{x}_\perp(\tau),\tau)\;\;,
\label{dee1}
\end{equation}
where $C_2$ is the color Casimir of the jet parton and $\kappa(E)$
is a slowly varying function of the jet energy \cite{Gyulassy:2003mc}. 
The azimuthal angle sensitive
escape time, $L(\phi)$, depends on the initial production point and
direction of propagation relative to the elliptic flow axis of the 
QGP~\cite{Gyulassy:2000gk}. For a longitudinal expanding
QGP, isentropic perfect fluid flow
implies that  $\tau \rho(\tau) \approx (1/A_\perp) dN_g/dy$ is fixed by the initial
gluon rapidity  density.  In this case
\begin{equation}
\frac{\Delta E(\phi)}{E} \propto C_2 \frac{L(\phi)}{A_\perp} \frac{dN_g}{dy}\propto C_2 
N_{part}^{2/3} \frac{L(\phi)}{<L>} \;\;.
\label{dee2}
\end{equation}
Therefore, gluon jets are expected to lose about $9/4$ more energy than quarks.
In addition, since the produced gluon density scales as the number of wounded participating
nucleons at a given impact parameter, (\ref{dee2}) predicts a particular centrality 
and azimuthal dependence of the energy loss. Detailed numerical studies show that 
the actual GLV energy loss  
can account qualitatively for the saturation of 
$v_2(p_T>2)$~\cite{Gyulassy:2000gk}, the unexpected $p_T$ 
independence~\cite{Vitev:2002pf} of the quenching pattern, the centrality 
dependence~\cite{Adler:2003qi} of the suppression factor~\cite{mgcipanp,agv}, 
and the rapidity dependence~\cite{Arsene:2003yk} of 
$R_{AA}(\eta,p_T)$~\cite{Hirano:2003yp,agv} for pions and high 
$p_T > 6$~GeV inclusive charged hadrons at RHIC.

To further illustrate qualitatively how (\ref{dee2}) 
influences the quench pattern, consider a  simplified initial
jet distribution rate, $d^2N/d^2p_0 = c p_0^{-n}$ where $n\sim 7$.
In applications these are of course calculated numerically.
After passing through the QGP, the final jet $p_T=p_f=p_0(1-\epsilon)$. 
The average over fluctuations
constrains $\langle \epsilon\rangle
=\Delta E(\phi)/E$. The quenched jet distribution is $d^2N/d^2p_T =(d^2 p_0/d^2p_f) 
d^2N/d^2p_0 \approx
(1-\epsilon(\phi))^{n-2} \; c p_f^{-n}$. The calculated hadron inclusive distribution
is obtained by folding the quench jet distribution 
over the  fragmentation function
$D(z=p_h/p_f,Q^2=p_f^2)$. However, in this illustrative example the fragmentation
function dependence drops out, and the nuclear
modification factor, $R_{AA}(p_h, \phi,N_{part})=dN_h(\epsilon)/{dN_h(0)}$ reduces to 
\begin{equation}
R_{AA}  
= \langle (1-\epsilon(\phi))^{n-2}\rangle\approx  
\left\langle \left(1- \epsilon_c 
\frac{L(\phi)}{\langle L\rangle} \left(\frac{N_{part}}{2A} \right)^{2/3} \right)^{n-2} 
\right\rangle
\label{raa} 
\end{equation}
The average over $\epsilon$ 
takes into account fluctuations of the radiative energy loss. 
The resulting $R_{AA}$ is independent of $p_T$ in this approximation. This was also found in 
the detailed numerical work in \cite{Vitev:2002pf}.
In central collisions, $N_{part}\approx 2 A$ and $L(\phi)\approx {\langle L\rangle}$,
and the magnitude of quenching is fixed by $\epsilon_c\propto
dN_g/dy$. The centrality and azimuthal dependence
for non-central collisions follows without additional calculations.

\subsection{\protect{$I_{AA}$} and Di-Jet Tomography}  

\begin{figure}[htp]
\centering
\vspace*{0.4cm}\hfill
\includegraphics[height=0.33\textheight,width=0.47\textwidth,clip] {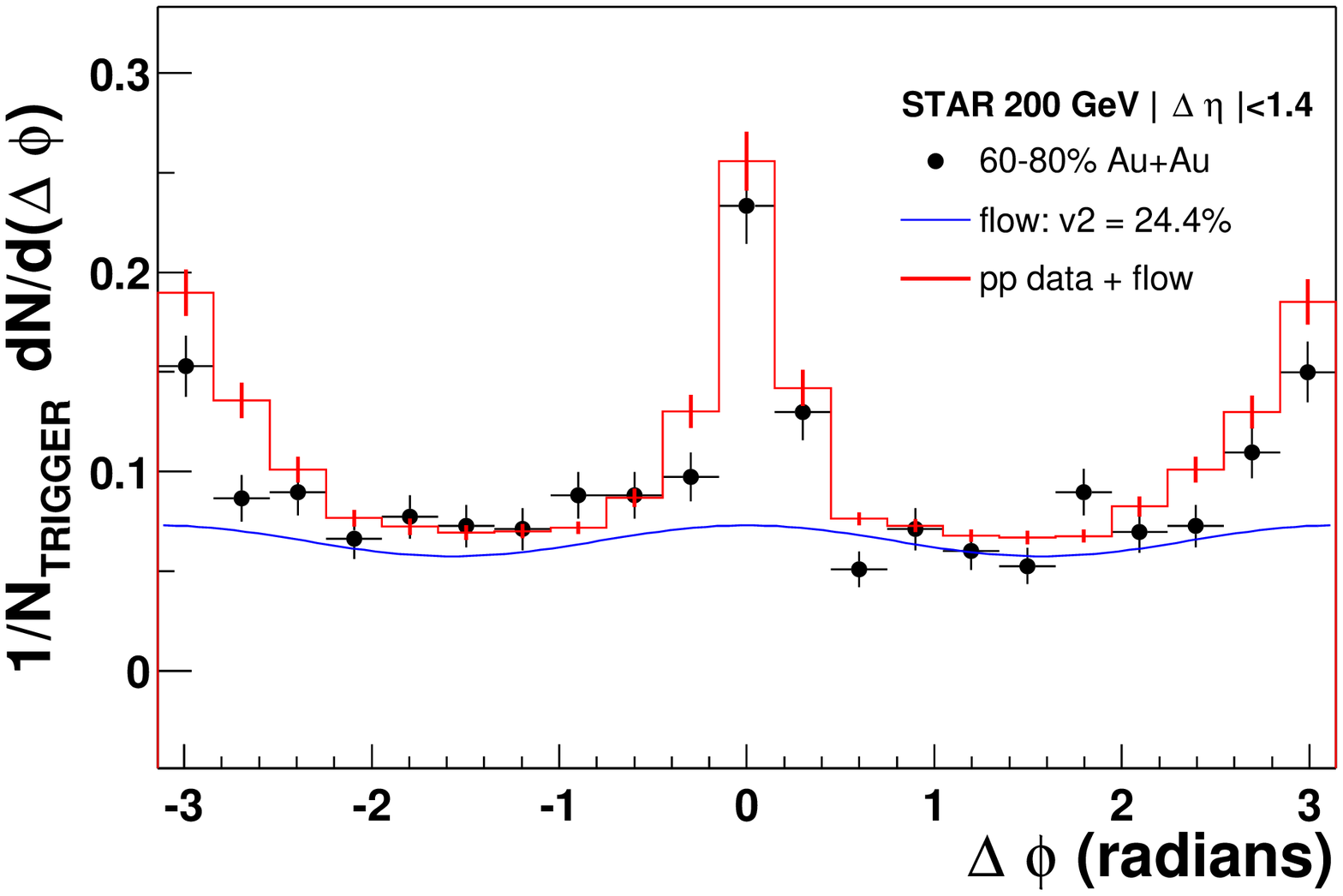}
\includegraphics[height=0.33\textheight,width=0.47\textwidth,clip] {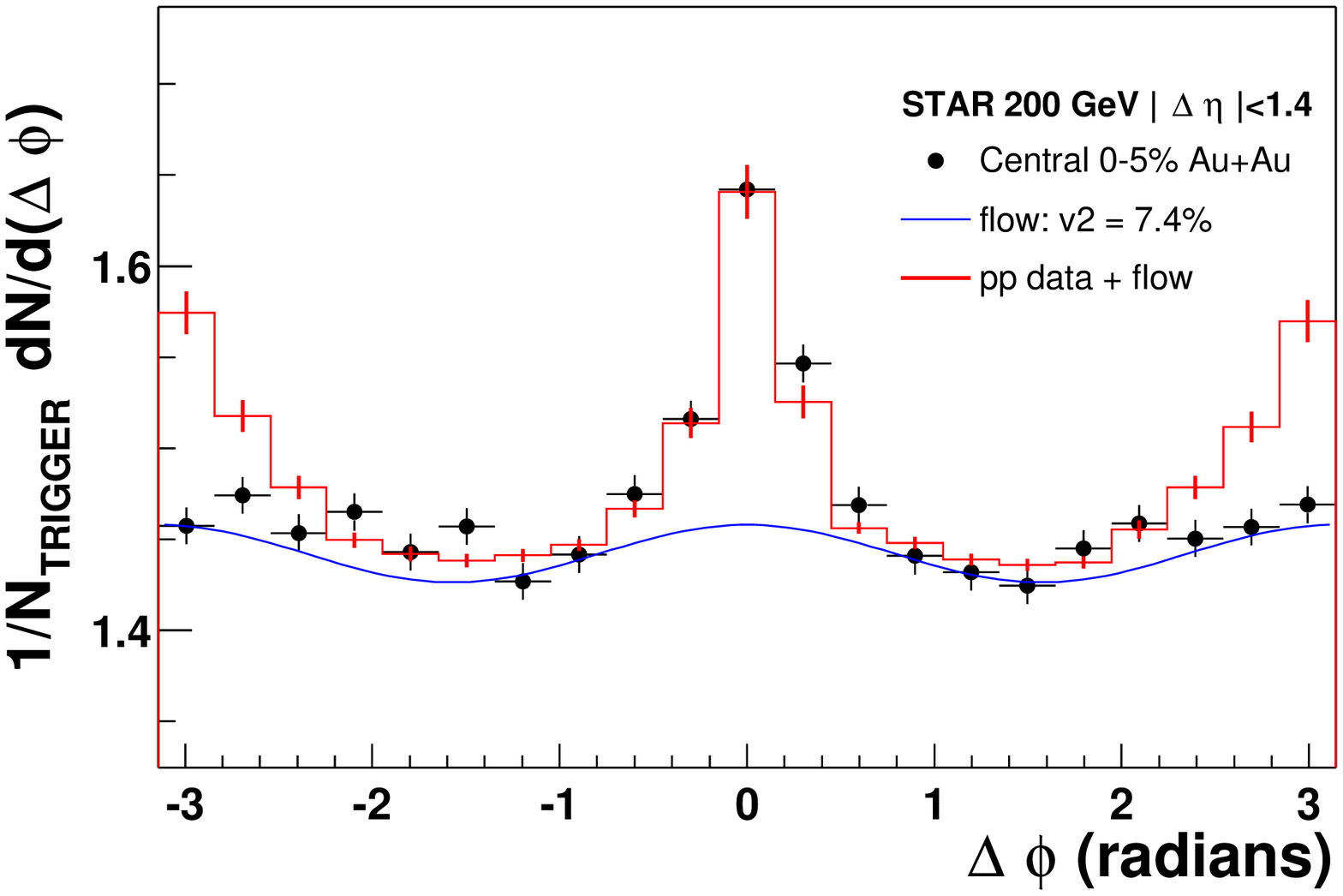}
\caption{Monojets at RHIC
 from STAR~\protect{\cite{Adler:2002tq,Hardtke:2002ph,Jacobs:2003bx}}. 
Strongly correlated back-to-back di-jet production 
in $pp$ and peripheral $AuAu$ left side is compared to mono-jet production
discovered in central $AuAu$.}
\label{monojet}
\end{figure}

Measurements of near side and away side azimuthal angle correlations of
di-jet fragments provide the opportunity to probe
the evolution of the QGP color charge density in even more detail.
Fig.(\ref{monojet}) show the discovery~\cite{Jacobs:2003bx,Adler:2002tq,Hardtke:2002ph} 
of mono-jet production~\cite{Gyulassy:1990bh} in central 
collisions at RHIC.
In peripheral collisions, the distribution $dN/ d\Delta \phi$ of the azimuthal
distribution of $p_T\sim 2$ GeV hadrons relative to a tagged $p_T\sim 4$ GeV
leading jet fragment shows the same near side and away side back-to-back jetty correlations
as measured in $p+p$. This is a direct proof that the kinematic range studied
tests the physics of pQCD binary parton collision processes. 
For central collisions, on the other hand, away side jet correlations are almost completely
suppressed.

The quantitative measure of the nuclear modification of di-jet correlations in $A+B$
reactions at a given $\sqrt{s}$ is given
by a formidable multi-variable function  
\begin{equation}
C_{AB}(y_1,p_{T1},\phi_1, y_2, p_{T2}, \phi_2; b, \Phi_b, h_1, h_2) \;\;,
\label{cab}
\end{equation}
where  $(y_1,p_{T1},\phi_1)$ is the trigger particle  of flavor $h_1$
and the  $(y_2,p_{T2},\phi_2)$ is an associated particle  of flavor $h_2$
for collisions at an impact parameter $b$ with a collective flow axis,
the reaction plane, fixing the azimuthal angle $\Phi_b$. 
Obviously $C_{AB}$ is a very powerful
microscope to study the modification of short wavelength correlations in 
the strongly interacting QGP.

The published data are as yet limited to $y_1\approx y_2\approx 0$, broad $p_T$
cuts: $p_{T1} > 4$ GeV and $p_T\sim 2$ GeV, 
two bins of $\phi_1-\phi_2$, and of course averaged over $\Phi_b$.
The measured modification of di-jet correlations is obtained by subtracting out 
the correlations due to bulk elliptic flow  via the di-jet measure
\begin{eqnarray}
I_{AA}&=& \int_{\Delta_{-}}^{\Delta_{+}} d(\phi_1-\phi_2)
\left\{N(\phi_1-\phi_2) \nonumber \right. \\
&& \left. \quad \quad - N_B(1+ 2v_2(p_1)v_2(p_2)\cos(2(\phi_1-\phi_2)) 
\right\} \;\;,
\end{eqnarray}
where the number of triggered pairs $N(\phi_1-\phi_2)$ is normalized relative to the 
expected number based on $p+p$ measurements in the same $[\Delta_{-},\Delta_{+}]$
relative azimuthal angle range. Wang~\cite{Wang:2003mm, Wang:2003aw} 
has analyzed the centrality, $N_{part}$, dependence of $I_{AA}$ as well as $R_{AA}$
and showed that both can be understood from the same pQCD
energy loss formalism. This provides another critical consistency check
of jet tomography at RHIC.

Additional preliminary data from STAR presented by K. Filimonov at Quark Matter 2004
showed the first direct evidence that back-to-back dijet quenching has a 
 distinctive dependence on the azimuthal orientation of the jets relative to the reaction
plane as expected from the obvious generalization of
Eq.(\ref{raa})
\begin{equation}
I_{AA}(\Delta \phi= \pi, \Phi_b)
\approx 
\langle\{(1- k_b L_1(\vec{r}_0,\hat{p}_{T}))(1- k_b L_2(\vec{r}_0,-\hat{p}_{T}))\}^{n-2}
\rangle 
\label{iaa} 
\end{equation}
where $\vec{r}_0$ is the initial transverse production point of the 
approximately back-to-back dijet moving in directions
$\hat{p}_{T}$ relative to the reaction plane $\Phi_b$.
Here $k_b$ the effective fractional energy loss per unit 
length in  the QGP produced at impact parameter $b$. 
The high $p_{T1}$ trigger naturally biases $\vec{r}_0$ to be near the surface 
and $\hat{p}_T$ is biased toward the outward normal direction.
This means that on the average, $L_1 \ll L_2$, and the away side fragments should be
strongly suppressed in the most central collisions, while the near side fragment
correlations should be similar to that seen in $pp$. However, in non-central
minimum biased events,  Eq.(\ref{iaa}) naturally predicts the away side fragments
are less quenched when the trigger hadron lies in the reaction plane 
than perpendicular to it, as observed by STAR.

\subsection{The Empirical {\bf pQCD} Line of Evidence}
Single and dijet data for pions above 3 GeV and protons above 6 GeV 
provide conclusive evidence that the QGP matter is partially opaque
to short wavelength probes with a quenching pattern as predicted by 
the {\bf pQCD} radiative energy loss: 
\begin{equation}
{\rm \bf pQCD}= R_{AuAu}(p_T,\phi, N_{part}) 
+ I_{AuAu}(\phi_1-\phi_2; b, \Phi_b) \;\;.
\label{pqcd}
\end{equation}  
With the vastly increased statistics from the current RHIC RUN 4, 
the tests of consistency of the theory
will be further extended to $p_T \sim 20$ GeV, and  $I_{AA}$ and $C_{AA}$
will attain ever greater resolving power. 
In addition,  heavy quark tomography~\cite{Dokshitzer:2001zm,
Djordjevic:2003zk,Batsouli:2002qf}
will provide new tests of the theory.

\section{The {\bf dAu} Control}

Only one year ago~\cite{transdyn03} the interpretation of high $p_T$ suppression
was under intense debate because it was not yet clear how much of the quenching
was due to initial state saturation (shadowing) of the gluon distributions and how much
due to jet quenching discussed in the previous section.
There was only one way to find out - eliminate the QGP final state interactions
by substituting a  Deuterium beam for one of the two heavy nuclei. In fact,
it was long ago anticipated~\cite{Wang:1992xy} 
 that such a control test would be needed to isolate
the unknown nuclear gluon shadowing contribution to the A+A quench pattern.
In addition $D+Au$ was required to test predictions of possible  
initial state Cronin multiple 
interactions~\cite{Wang:1998ww,Wang:1996yf,Vitev:2003xu,Qiu:2003vd}.
In contrast, one  model of 
CGC~\cite{Kharzeev:2002pc} predicted a 30\% suppression in central D+Au.

The data~\cite{Adler:2003ii,Adams:2003im,Arsene:2003yk,Back:2003ns} conclusively
rule out large initial shadowing as the cause of the $x_{BJ} > 0.01$
quenching in Au+Au and establish the empirical control analog of Eq.(\ref{pqcd})
\begin{equation}
{\rm \bf dA}= R_{DAu}(p_T,\phi, N_{part}) + I_{DAu}(\phi_1-\phi_2,b) \;\;.
\label{da}
\end{equation} 

\begin{figure}[t]
\centering
\vspace*{-0.6cm}
\includegraphics[height=0.30\textheight,width=1.0\textwidth,clip]{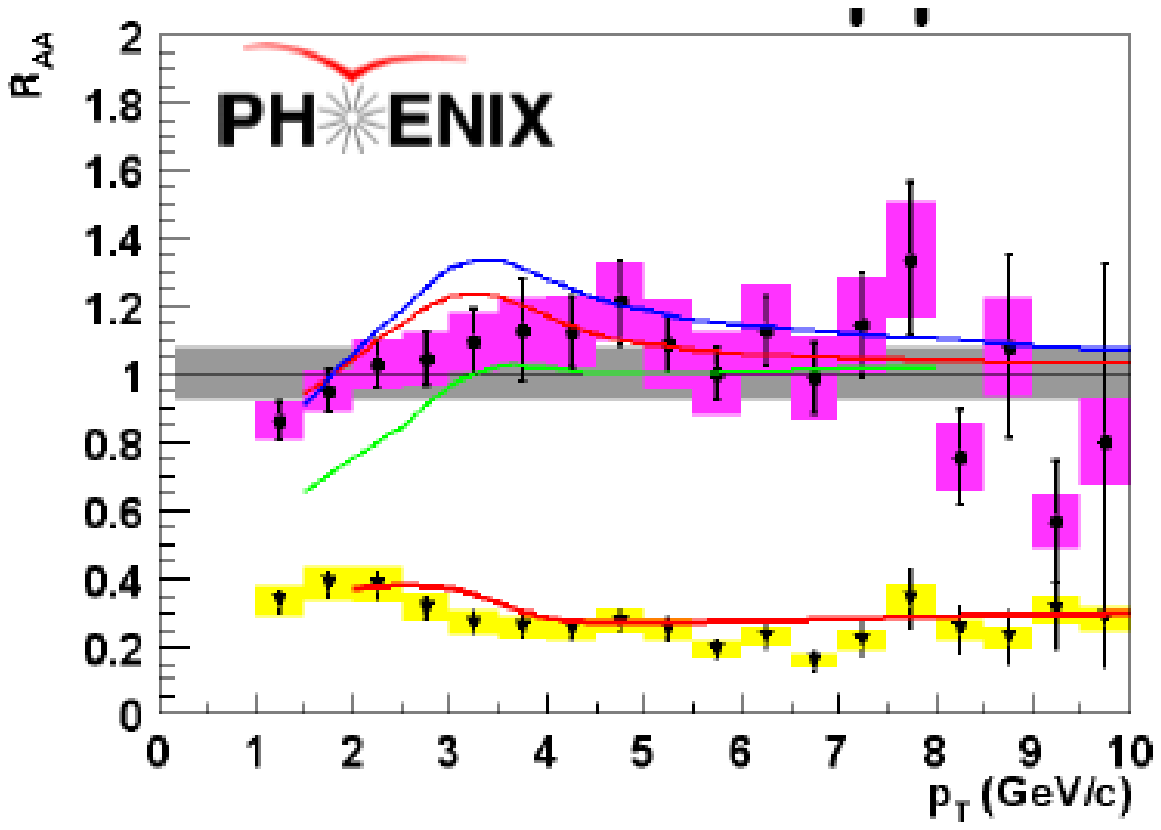}
\vspace*{-0.7cm}
\includegraphics[height=0.30\textheight,width=1.0\textwidth,clip]{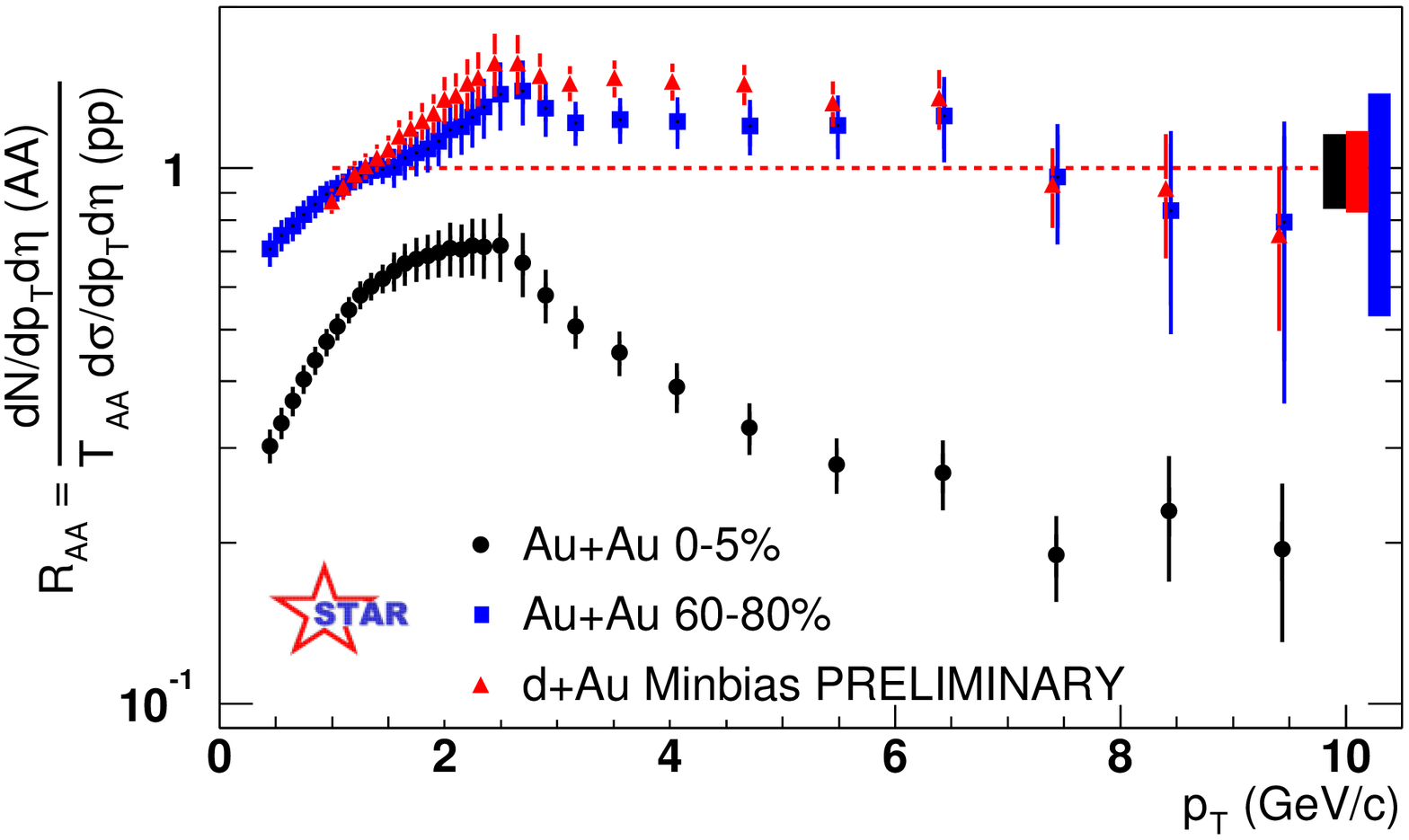}
\caption{The {\bf dA} control: PHENIX~\protect{\cite{Adler:2003ii}} $\pi^0$ 
and STAR~\protect{\cite{Adams:2003im}} $h^{\pm}$
data compare $R_{DAu}$ to $R_{AuAu}$. These and BRAHMS~\protect{\cite{Arsene:2003yk}}
 and PHOBOS~\protect{\cite{Back:2003ns}} data prove that jet quenching in $Au+Au$
must be due to final state interactions. Curves for $\pi^0$ show predictions from 
\protect{\cite{Vitev:2002pf}} for $AuAu$ and from
\protect{\cite{Vitev:2003xu}} $DAu$. The curves for $DAu$ show the interplay
between different gluon shadow parameterizations (EKS, none, HIJING) and Cronin enhancement
and are  similar to 
predictions in \protect{\cite{Wang:1998ww,Wang:1996yf,Vitev:2003xu}}.
In lower panel, the unquenching of charged hadrons 
is also seen in $D+Au$ relative to $Au+Au$ at high $p_T$.}
\label{rdaudata}
\end{figure}

\subsection{The ``Return of the Jeti''}
The $I_{DAu}$ measurement from STAR~\cite{Adams:2003im} is the check mate!
\begin{figure}[h]
\centering
\includegraphics[height=0.35\textheight,width=1.0\textwidth,clip]{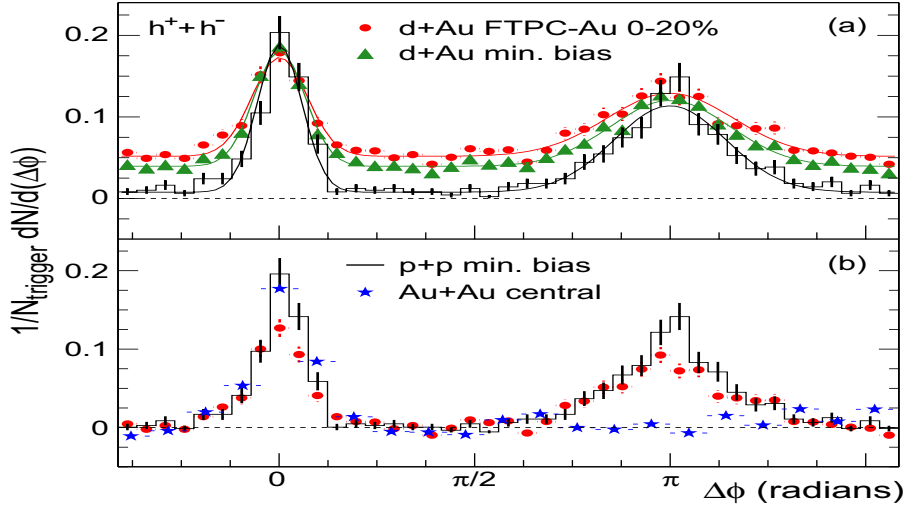}
\caption{The {\bf dA} ``Return of the Jeti'': 
Dijet fragment azimuthal correlations from STAR~\protect{\cite{Adams:2003im}} 
in $DAu$ are unquenched relative to the mono jet correlation
observed in central ${AuAu}$. }
\label{jeti}
\end{figure}
The return of back-to-back jet correlation in $D+Au$ to the level observed in $pp$
is seen in Fig.\ref{jeti}. The data rule out CGC gluon fusion models
that predict  mono-jets~\cite{Zoller:2003zs} correlations 
in the $x_{BJ}>0.01$ region.
These $D+Au$ data support the conclusion~\cite{Wang:2003mm,Wang:2003aw}
that the observed jet quenching in $AuAu$ is due to parton energy loss.

Another independent check of the strikingly different nature 
of the nuclear matter created in $Au+Au$ versus $D+Au$ 
is provided by the width 
of the away side correlation 
function~\cite{Qiu:2003pm}. The transport properties of cold nuclear matter
extracted from the Cronin enhancement effect~\cite{Vitev:2003xu} have only 
a small effect on the measured dijet acoplanarity. Preliminary PHENIX data 
presented by J.~Rak at Quark Matter 2004 confirm the qualitative 
similarity of $p+p$ and $D+Au$ for $p_T > 2$~GeV, 
but at the same time demonstrate  a strong  quantifiable increase in acoplanarity 
in central $Au+Au$ consistent with multiple semi-hard 
scattering~\cite{Gyulassy:2002yv} in a dense QGP.

\section{Conclusions}
The three lines of evidence have converged from the four RHIC experiments.
The empirical QGP that has been found  at RHIC via the 
combination of Eqs.(\ref{qgpdef},\ref{Pdef2},\ref{pqcd},\ref{da}).
This QGP is, however, not the weakly interacting~\cite{Collins:1974ky}, 
color-dielectric ``wQGP'', that we have searched for.
Because of its near perfect fluid long wavelength properties, it must be very 
strongly coupled  at least up to several times $T_c$. Symbolically, we should denote
\footnote{I thank T.D.Lee for 
discussions and suggesting the sQGP designation of 
the matter discovered at RHIC to distinguish it from weakly interacting,
Debye screened plasmas,  wQGP, which may only exist at temperatures
$T\gg T_c$.}
the  empirical QGP at RHIC by ``{\bf sQGP}'' to emphasize its special
properties. The {\bf sQGP} is not only a near perfect fluid but it
also retains part of its QCD asymptotic freedom character 
through its highly suppressed, but power law, short wavelength
spectrum.

In summary, the {\bf sQGP} found  RHIC was seen through the following
three convergent lines of evidence 
 \begin{eqnarray}
{\rm \bf sQGP}&=& {\rm \bf P_{QCD} + pQCD + dA} \nonumber \\
 &=& \{v_2(p_T; \pi,K,p,\Lambda) + v_1(y) + \;{ \rm\bf c.c.} \}\nonumber \\
   &\;& + \{R_{AuAu}(p_T,\phi, N_{part}) + I_{AuAu}(\phi_1-\phi_2; b, \Phi_b)\} \nonumber \\
   &\;& + \{ R_{DAu}(p_T,\phi, N_{part}) + I_{DAu}(\phi_1-\phi_2, b)\}
\;\; .
\label{qed}
\end{eqnarray}
Other surprising properties are already known.
The anomalous $p_T=2-5$ GeV baryon/meson 
ratios~\cite{Sorensen:2003kp,Adler:2003cb} ,
noted in connection
with Fig.3 and also seen indirectly in Fig.8,
already point to unexpected novel baryon number physics. Current speculations  
center around possible gluonic baryon junction dynamics~\cite{Kharzeev:1996sq}, 
and possible  multi-quark quark coalescence mechanisms~\cite{Csizmadia:1998vp}.
The baryon number transport properties in the sQGP  
will certainly teach us new physics.

The experimental task of mapping out the novel properties of sQGP 
has only begun. It is important to concentrate, however, on those observables
which are least distorted or ``polluted'' by uninteresting 
hadronic final interactions.
For example, the severity of the HBT puzzle depends on
which hadronic transport model is used. Pure hydrodynamics 
with late freeze-out times fails badly  to reproduce  final state soft pion correlations.
Hybrid hydro+RQMD hadronic cascade does somewhat better~\cite{Soff:2000eh}, but 
there is at least one transport model~\cite{Lin:2002gc} that reproduced the data. 
Non-equilibrium hadron resonance transport dynamics are unfortunately still not well
enough understood at any energy to allow definitive conclusions to be drawn.

Thermal direct photons have yet to be measured, but it is already
known that the pre-equilibrium~\cite{Wang:1998ww} and hadronic final
state contributions in the few GeV $p_T$ range will produce large
backgrounds on top of the thermal component. These must be
deconvoluted if thermal photons are to serve as a sQGP thermometer.
However, even the theoretical thermal photon rates are still under
debate~\cite{Gelis:2002yw}.

The $J/\psi$ suppression discovery at SPS was originally attributed to 
Debye screening of $c\bar{c}$ in a  wQGP paradigm.
Recent lattice QCD results
now indicate that heavy quark correlations persist perhaps up to $~2T_c$.
This is another indication for the strongly coupled nature of 
{\bf sQCD}. The suppression of $J/\psi$ in $AA$ is also strongly influenced by 
initial state and final state hadronic (comover) effects.
It will be no easier to  deconvolute these competing effects at RHIC.
These observables of course need to be measured at RHIC, but one should not
expect an easy interpretation.

I believe that the most promising direction of future experiments at RHIC
will be precision measurements of short wavelength 
($p_T>2$ GeV) correlators $C_{AB}$ illustrated in Eq.(\ref{cab}). These
are very  powerful six dimensional microscopes with four discrete
$(h_1,h_2, A, B)$ and two continuously adjustable geometric 
$(b,\Phi_b)$ experimental knobs in addition to the beam 
energy $\sqrt{s}=20-200$~AGeV. One of the important
correlators will be that of direct photon tagged
jets~\cite{Wang:1996yh}. Another will be open charm and possibly bottom
quark tomographic probes. The available experimental knobs
have hardly been varied yet. Much remains to be done to map out 
to clarify the properties of the {\bf sQGP} found  at RHIC.

\begin{acknowledgments}
This work was supported by the Director, Office of Energy
Research, Office of High Energy and Nuclear Physics,
Division of Nuclear Physics,
of the U.S. Department of Energy
under Contract DE-FG-02-93ER-40764.
I also gratefully acknowledge partial support 
from an Alexander von Humboldt-Stiftung Foundation for 
continuation of collaborative
work at the Institut f\"ur Theoretische Physik, Frankfurt.
\end{acknowledgments}

\bibliographystyle{amsunsrt}
\chapbblname{sample} 
\chapbibliography{sample} 


\end{document}